\documentclass[UTF8,twocolumn]{ctexart}
\usepackage{amsmath}
\usepackage{amssymb}
\usepackage{booktabs}
\usepackage{tabularx}
\usepackage{float}
\usepackage{placeins}
\usepackage{capt-of}
\usepackage{enumitem}
\usepackage{xcolor}
\usepackage{tikz}

\usepackage{hyperref}
\usepackage{xurl} 
\hypersetup{pdfborder={0 0 0}}

\usetikzlibrary{arrows.meta,positioning,fit}
\usepackage{graphicx}
\usepackage[
    a4paper,
    top=2.2cm,
    bottom=2.2cm,
    left=1.8cm,
    right=1.8cm
]{geometry}

\usepackage{titlesec}
\renewcommand{\thesection}{\arabic{section}}

\usepackage{cuted}
\usepackage{listings}
\lstdefinestyle{cqlibpython}{
  language=Python,
  basicstyle=\ttfamily\scriptsize,
  keywordstyle=\bfseries,
  commentstyle=\itshape,
  showstringspaces=false,
  breaklines=true,
  columns=fullflexible,
  keepspaces=true,
  frame=single,
  framesep=4pt
}

\title{\textit{\textbf{Cqlib}: A High-Performance Software Stack for Quantum Computing}}

\author{
\textit{Tianyan Quantum Group}\\
China Telecom Quantum Information Technology Group Co., Ltd., Hefei 230031, China
}

\date{September 2026}

\begin{document}

\twocolumn[
\begin{@twocolumnfalse}

\maketitle

\begin{abstract}

This paper presents Cqlib, a high-performance software stack for quantum computing built around a unified quantum circuit model. Its native Rust core provides efficient implementations of circuit processing, quantum-state and operator computation, compilation, and optimization, while Python and C interfaces support algorithm development and system integration. Cqlib covers quantum program construction, simulation, analysis, compilation, device adaptation, and result processing under consistent program semantics. 
Through OpenQASM and the Quantum Control Instruction Set (QCIS), Cqlib supports program representation and hardware-oriented execution. Cqlib further provides backend interfaces for device execution, with Cqlib-Tianyan serving as one implementation for QCIS-compatible quantum hardware. Evaluations using parameterized compilation tasks from Benchpress and related tests show that Cqlib achieves good functional coverage, compilation correctness, and execution efficiency within the evaluated task scope. 
We develop Cqlib-Algorithm as a modular library of reusable operators and composable components for widely used quantum algorithms, including quantum search, phase estimation, and variational algorithms, with additional support for error mitigation. To complement these building blocks with domain-specific workflows, we further develop Cqlib-VQE, Cqlib-QAOA, and Cqlib-QML for application-level studies in quantum chemistry, combinatorial optimization, and quantum machine learning, respectively. 
Cqlib also integrates source-grounded Skills and MCP-based tools to support agent-assisted program generation, backend execution, and result-guided workflow refinement. Overall, Cqlib provides a unified and extensible software foundation that connects quantum algorithm development, simulation, compilation, and execution on QCIS-compatible quantum hardware.

\setlength{\parskip}{\baselineskip} 

\noindent Cqlib website:~\href{https://qc.zdxlz.com/cqlib}{\textcolor{blue}{\nolinkurl{https://qc.zdxlz.com/cqlib}}}

\end{abstract}

\vspace{1em}

\end{@twocolumnfalse}
]

\section{Introduction}

Quantum computing is gradually progressing from small-scale proof-of-principle demonstrations toward systematic algorithm research based on programmable quantum processors. As qubit counts, gate fidelities, and quantum--classical coordination capabilities continue to improve, quantum programs are increasing in scale and complexity, placing greater demands on program representation, compilation optimization, hardware adaptation, and execution efficiency~\cite{preskill2018nisq,cerezo2021vqa,javadiabhari2024qiskit}. This paper presents Cqlib 2.0 (China Quantum Library), a high-performance software stack for quantum computing. 

Cqlib is designed around a unified core and a layered architecture with clearly defined component responsibilities and extensible frontend and backend interfaces.
It uses quantum circuits as a unified program representation and integrates multilingual programming interfaces, quantum information processing, target-device compilation, local simulation, backend execution, and result processing under consistent program semantics~\cite{cross2017openqasm,cross2022openqasm3}.

To support efficient implementation and cross-platform extension of this software architecture, Cqlib adopts a high-performance Rust kernel and coordinated multilingual interfaces. Under a unified data model and consistent execution semantics, this architecture combines high-level program representation with low-level computational efficiency. For frequently executed tasks such as parameterized circuit processing, quantum-state evolution, and operator evaluation, Cqlib executes performance-critical operations in native numerical kernels and employs parallelism where beneficial, reducing language-boundary overhead and improving execution efficiency. For physical quantum devices, the compilation process further incorporates device topologies, target gate sets, and hardware constraints to adapt logical circuits into device-executable programs. Cqlib thereby connects algorithm development, high-performance computation, and device execution under unified program semantics, providing a stable and extensible software foundation for moving quantum algorithms from prototype validation to engineering implementation.

We further evaluate Cqlib using the Benchpress benchmark~\cite{nation2025benchpress} and multilingual compilation tests. The evaluation examines functional coverage, compilation correctness, and cross-language performance. The results show that Cqlib satisfies the applicable structural and semantic validation criteria across the compilation tasks evaluated in this work. Different language interfaces also maintain comparable compilation performance while sharing the same Rust core. Beyond compilation, Cqlib supports algorithm development through Cqlib-Algorithm, which provides common protocols, reusable quantum operators, composable components, and end-to-end algorithm workflows. The domain-specific frameworks Cqlib-VQE, Cqlib-QAOA, and Cqlib-QML demonstrate the extensibility of Cqlib for applications in quantum chemistry, combinatorial optimization, and quantum machine learning, respectively.

Relationship with previous work: Our previous work on the Tianyan Quantum Cloud Platform~\cite{tianyan2025cloud} focused on cloud-accessible quantum hardware and the experimental demonstration of the quantum computational advantage.
Cqlib was introduced there primarily as the software interface supporting experiments conducted through the Tianyan cloud platform, including circuit construction, compilation, task execution, and result processing. 
The present manuscript corresponds to Cqlib 2.0, a substantially upgraded version of the software system, and studies Cqlib itself in terms of its architecture, implementation, and evaluation.
Cqlib is designed as a general and extensible quantum software framework rather than a Tianyan-specific interface, with Cqlib-Tianyan serving as one concrete backend implementation.
The study covers its unified quantum-program representation, high-performance Rust core, compiler architecture, multilingual interfaces, systematic benchmarking, application frameworks, and agent-assisted programming capabilities. The hardware characterization and random-circuit-sampling experiments reported in Ref.~\cite{tianyan2025cloud} are not contributions of the present work.

The remainder of this paper is organized as follows. Section~\ref{sec:motivation} presents the motivation and design objectives of Cqlib. Section~\ref{sec:cqlib-Architecture} introduces the overall architecture. The core design of the high-performance computing kernel and quantum compiler are described in Section~\ref{sec:core-design}. The Benchpress benchmark and multilingual compilation tests are presented in Section~\ref{sec:benchpress-benchmark-evaluation}. Then, Section~\ref{sec:quantum-computing-frameworks-based-on-cqlib} introduces the Algorithm library, VQE, QAOA, and QML frameworks built on Cqlib. Section~\ref{sec:cqlib-skills} presents the integration of Cqlib Skills and MCP for agent-assisted quantum programming and execution. Finally, Section~\ref{sec:related-work} discusses related work and Section~\ref{sec:conclusion-and-outlook} concludes the paper and outlines future directions.

\section{Motivation}
\label{sec:motivation}

Quantum computing applications are gradually expanding from algorithm validation to experimental studies on physical devices. A quantum program passes through problem formulation, circuit construction, simulation-based validation, compilation, and device execution. The representations and interfaces used between these stages affect program consistency, execution efficiency, and experimental reproducibility. Cqlib is designed to establish a unified software foundation that reduces repeated conversions among these stages while supporting algorithm development,
improving computational efficiency, and facilitating hardware adaptation.

\subsection{Unified Program Processing Requirements}

A quantum application undergoes multiple program transformations from an algorithm description to hardware execution. A high-level algorithm is first implemented as an appropriate quantum circuit. The circuit is then transformed through gate decomposition, circuit optimization, qubit mapping, and routing into a program accepted by the target device. Execution results must subsequently be returned to classical programs for statistical analysis, parameter updates, or generation of the next quantum task. For hybrid quantum--classical applications such as variational quantum algorithms and QML, this process is repeated frequently.

If different processing stages use different circuit representations and parameter conventions, developers must perform additional data conversions. These differences may also lead to inconsistent program behavior, omitted device constraints, and difficulties in diagnosing errors. A common quantum circuit representation is therefore needed to connect the processing stages and define the inputs, outputs, and responsibilities of individual modules. Cqlib organizes circuit construction, compilation, and execution as a continuous program-processing workflow, allowing quantum programs to undergo local validation and device execution while maintaining a consistent representation.

\subsection{Multilingual Interfaces and Performance}

Quantum software must support rapid algorithm prototyping and performance-sensitive workloads while enabling integration with existing engineering systems. Python is suitable for algorithm development, data analysis, and interactive validation, while C interfaces enable integration with C/C++ applications and high-performance computing environments. Frequently executed operations, including circuit processing, compilation, and quantum information processing, require a stable and controllable low-level implementation. If different languages maintain separate circuit objects, parameter systems, and compilation logic, interface convenience comes at the cost of semantic divergence, data copying, and additional maintenance.

Cqlib implements circuit definitions, parameter handling, quantum information processing, and compiler infrastructure in a shared Rust core. Python and C interfaces are built on top of this core implementation. This architecture is designed to maintain consistent program semantics and numerical accuracy across language environments while moving computationally intensive operations to native code. It thereby connects efficient algorithm development, low-level execution performance, and software reuse.

\subsection{QCIS and Device Compilation}

Quantum algorithms are commonly represented as logical circuits that are independent of specific hardware. Quantum processors differ in their available qubits, coupling topologies, native gate sets, and instruction formats. If device constraints are processed only before task submission, circuit optimization remains separated from actual hardware conditions, and developers must repeatedly convert among logical circuits, physical circuits, and device instructions. These conversions increase adaptation costs, make semantic consistency more difficult to maintain, and complicate the analysis of compilation overhead and experimental results.

The Quantum Control Instruction Set (QCIS) provides an instruction-level interface for controlling superconducting quantum computing hardware ~\cite{wu2021strong}. To support QCIS-compatible quantum devices, Cqlib incorporates device constraints into a unified quantum program transformation process, allowing logical circuits to be converted into device-executable programs while preserving their intended computation. 
A well-defined interface links the common circuit representation to QCIS instructions, reducing the development and validation effort required to adapt quantum algorithms to target hardware.

\begin{figure*}[!t]
  \centering
  \includegraphics[
    width=1\textwidth,
    keepaspectratio
  ]{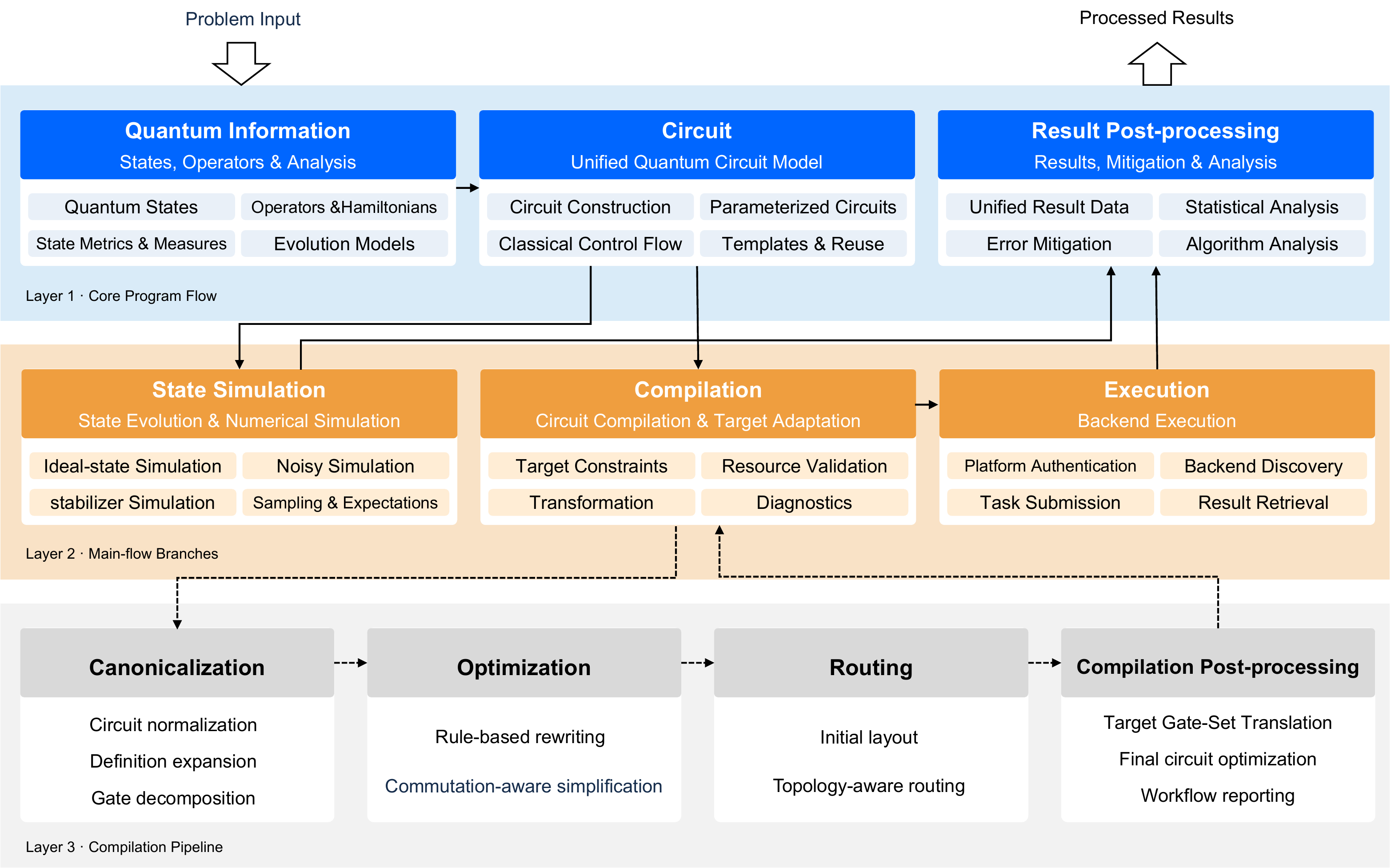}
  \caption{Main Cqlib workflow and compilation pipeline}
  \label{fig:Cqlib-overall-architecture}
\end{figure*}

\subsection{Design Objectives}

Cqlib is not intended simply to combine circuit tools, compiler components, and cloud interfaces. Its goal is to establish a unified software foundation that preserves quantum program semantics throughout program processing. The main design objectives are as follows:

\begin{itemize}[leftmargin=1.5em]
  \item \textbf{Unified Program Representation:}
  Shared circuit and intermediate representations connect circuit construction, parameterization, compilation, simulation, format conversion, and result processing. This design reduces semantic loss caused by conversions between modules.

  \item \textbf{Multilingual Support:}
  The Rust core provides performance-sensitive capabilities, while Python supports algorithm development and C supports system integration. All interfaces reuse the same core implementation to avoid divergence among language-specific implementations.

  \item \textbf{Support for QCIS-Compatible Quantum Devices:}
  Target gate sets, device topologies, and QCIS generation are incorporated into the compilation and execution workflow. Logical quantum programs can therefore be transformed into device-executable programs through a traceable compilation process.

  \item \textbf{Execution on Physical Devices:} Dedicated execution interfaces connect compiled circuits to hardware backends while keeping platform-specific service logic separate from the Cqlib core.
\end{itemize}

Cqlib therefore targets not isolated quantum circuit programming, but an extensible software infrastructure that connects quantum problem formulation, program optimization, and execution on physical devices.

\section{Cqlib Architecture}
\label{sec:cqlib-Architecture}

Cqlib supports the quantum programming workflow from problem formulation and circuit construction to simulation, compilation, execution, and result analysis. Its architecture follows three principles: a unified program model, reuse of core capabilities, and separation of execution interfaces. The \texttt{Circuit} object serves as the common representation of a quantum program. A Rust core implements circuit processing, quantum information processing, and compiler optimization. Python and C interfaces support algorithm development and system integration.

Cqlib focuses on reusable capabilities for quantum program processing. The quantum information module provides quantum states, operators, and numerical operations. The circuit module defines the common program representation. The state simulation module performs numerical simulation, while the compilation module transforms circuits for target devices. The execution module connects Cqlib to execution environments, and the postprocessing module prepares simulation or execution data for further analysis. 
For execution on the Tianyan quantum computing platform, dedicated backend interfaces submit tasks, query their status, and retrieve results. These interfaces maintain a clear boundary between general program processing and platform-specific execution.

Figure~\ref{fig:Cqlib-overall-architecture} shows the overall workflow and the relationships among the Cqlib modules. The \texttt{Circuit} representation connects all stages of program processing. The workflow then separates into two paths for state simulation and execution on physical devices. The figure also expands the normalization, optimization, routing, and postprocessing stages of the compilation pipeline.

This section describes the relationships among the components and the end-to-end data flow. It does not cover internal algorithms or implementation details. Section~\ref{sec:core-design} presents the main mechanisms for circuit representation, high-performance numerical kernels, gate decomposition, rule-based optimization, layout, and routing.

\subsection{System Components}

The main components of Cqlib are listed below. They operate on the same quantum program model. This subsection focuses on their functional boundaries.

\begin{itemize}[leftmargin=2em,itemsep=0.25em,topsep=0.4em]

  \item \textbf{Quantum Information Science (QIS):}
        Provides quantum states, operators, Hamiltonians, quantum state analysis, and quantum evolution. These capabilities support circuit construction, state simulation, and result analysis.

  \item \textbf{Unified Quantum Program Model (Circuit):}
        Represents quantum programs as \texttt{Circuit} objects. The model supports quantum gates, parameterized circuits, classical control flow, circuit templates, validation, and analysis.

  \item \textbf{State Simulation:}
        Supports ideal, noisy, and stabilizer simulation, as well as sampling and expectation-value evaluation. These functions enable local validation and numerical analysis of quantum programs.

  \item \textbf{Compilation:}
        Applies circuit normalization, optimization, layout, and routing under target-device constraints. It transforms a logical circuit into a representation that satisfies the requirements of the execution environment.

  \item \textbf{Execution:}
        Provides platform access, task submission, task-status management, and result retrieval. The \texttt{TianyanPlatform} and \texttt{TianyanBackend} interfaces adapt these functions to the Tianyan quantum computing platform.

  \item \textbf{Postprocessing:}
        Organizes simulation and hardware-execution results in a common format. It also supports statistical analysis, error mitigation, visualization, and subsequent algorithm analysis.
\end{itemize}

\subsection{Unified Program Processing Model}

Cqlib organizes the workflow around Circuit as its common quantum program representation, together with quantum-information objects for problem formulation and structured result types for analysis.
After circuit construction, a program enters either the state-simulation path or the compilation-and-execution path, depending on the execution target.

\begin{enumerate}[leftmargin=2.4em,itemsep=0.35em,topsep=0.4em]

  \item \textbf{Quantum Information and Problem Formulation:}
        The user constructs quantum states, Hamiltonians, observables, and other quantum information objects for the target problem. These objects define the data and computational objectives used during circuit construction.

  \item \textbf{Quantum Circuit Construction:}
        A \texttt{Circuit} object represents quantum gates, symbolic parameters, measurements, classical control flow, and reusable circuit structures. 
Subsequent simulation and compilation stages use the same Circuit representation. For device execution, the compiled circuit is converted to the instruction format required by the target backend.

  \item \textbf{State-simulation Path:}
        For algorithm validation and numerical analysis, a \texttt{Circuit} can be sent directly to the state-simulation module. Ideal, noisy, or stabilizer simulation performs state evolution, sampling, and expectation-value evaluation. The postprocessing module then processes the simulation results.

  \item \textbf{Compilation-and-Execution Path:}
        For execution on a target system, a \texttt{Circuit} first enters the compilation module. This module transforms the circuit and adapts it to target constraints. 
The execution module submits the task to the selected backend, monitors its status, and retrieves the results.
For the Tianyan platform, dedicated backend interfaces provide task submission, status queries, and result retrieval.

  \item \textbf{Result Processing and Output:}
        Results from state simulation and hardware execution enter the same postprocessing module. This module organizes measurement counts, probability distributions, and observable expectation values. It can also perform statistical analysis, error mitigation, and visualization to produce the final output.
\end{enumerate}

Cqlib therefore provides two main processing paths: local state simulation and compilation followed by execution on a target device. Both paths use the same \texttt{Circuit} representation and converge at a common postprocessing stage. This design maintains consistent program semantics and result interfaces across execution modes.

\subsection{Compilation Pipeline}

Within the main workflow, compilation consists of normalization, optimization, routing, and compilation postprocessing. Each stage operates on the same \texttt{Circuit} representation and progressively prepares, optimizes, and adapts the circuit for the target device.

\begin{center}
\small
\textbf{Normalization}
$\rightarrow$
\textbf{Optimization}
$\rightarrow$
\textbf{Routing}
\\[0.25em]
$\rightarrow$
\textbf{Compilation Postprocessing}
\end{center}

\textbf{Normalization} standardizes circuits, expands definitions, and decomposes operations into basic gates. It converts different circuit forms into a common internal representation. \textbf{Optimization} removes redundant operations through rule-based rewriting, circuit simplification, and commutation analysis. \textbf{Routing} maps logical qubits to physical resources and routes interactions according to the target topology. \textbf{Compilation Postprocessing} converts the circuit to the native gate set, performs final cleanup, and produces the compilation result.

This staged pipeline separates general circuit transformation, circuit optimization, and device-specific mapping. It preserves a common program representation while providing a consistent compilation workflow for different execution environments.

\section{Core Design}
\label{sec:core-design}

\subsection{High-Performance Core Design}

Cqlib combines unified computational abstractions, native execution kernels, and algorithm-level optimization. Its design targets the complete execution path rather than the cost of an individual quantum gate. This path includes circuit construction, parameter binding, state evolution, and expectation-value evaluation. The resulting foundation supports large circuit simulations and hybrid quantum--classical algorithms.

\subsubsection{Unified Computational Abstractions and System-Level Co-Design}

Cqlib preserves expressive high-level interfaces while moving performance-sensitive operations into a native core. The Rust core implements quantum circuits, parameter management, quantum states, operator application, and compiler transformations. Python and C/C++ bindings reuse the same data model and execution semantics. High-level languages handle problem specification, experiment orchestration, and result analysis. Computationally intensive operations remain in native code to avoid frequent language-boundary crossings during gate execution and amplitude updates.

This organization defines clear responsibilities across layers. Unified abstractions describe circuit structures and algorithm logic, while native kernels perform state evolution and operator evaluation. Language bindings handle only object conversion and task submission. The Python binding is implemented with PyO3, and the C interface exposes core capabilities through a stable foreign-function interface. Gate definitions, parameter rules, and result types therefore remain consistent across languages. For high-performance computing (HPC) systems, control software, and existing C/C++ applications, explicit creation and release functions manage object lifetimes. Python workflows invoke the kernel at the granularity of circuits, parameter sets, and execution requests. This design reduces overhead from object conversion, error wrapping, and reference counting.

Rust's ownership model, borrowing rules, and type system provide memory-safety guarantees for the native implementation. In safe Rust, these mechanisms constrain object lifetimes and mutable aliasing, helping prevent dangling references and data races. During parallel state-vector execution, mutable state data are partitioned into disjoint regions before being processed concurrently. Fallible operations return explicit errors for invalid parameters, dimension mismatches, and unsupported transformations, allowing errors to be propagated rather than silently ignored.

Variational quantum algorithms repeatedly update parameters while keeping the circuit structure fixed. Cqlib stores symbolic parameters in a circuit-level parameter table. 
For each scalar gate parameter, Cqlib stores either an index into the circuit-level parameter table or a fixed numerical value.

\begin{equation}
  p_{\mathrm{op}}\in
  \left\{\operatorname{Index}(k),\operatorname{Fixed}(v)\right\},
\end{equation}

Here, $k$ indexes the circuit-level parameter table, and $v$ is a floating-point value that has passed simplification and finiteness checks. Consider a circuit with $M$ parameterized operations and $K$ distinct symbolic expressions. Centralized storage reduces the number of symbolic objects from at most $M$ to $K$. During parameter binding, Cqlib first evaluates the parameter table and then resolves each gate parameter by index. This process separates symbolic evaluation from gate execution and allows the circuit topology to be reused across optimization iterations.

\subsubsection{Native Numerical Kernels and Architecture-Aware Optimization}

Cqlib improves performance through build-time optimization, optimized memory layouts, specialized kernels, SIMD (Single‑Instruction, Multiple‑Data) vectorization, and task-level parallelism.
Release builds enable link-time optimization and use a single code-generation unit. The compiler can therefore apply inlining, constant propagation, and unreachable-code elimination across module boundaries. These optimizations reduce the cost of short calls among gate operations, parameter resolution, and state updates while preserving modular abstractions. Symbol stripping further removes nonexecutable information from the final binary.

For an $n$-qubit pure state, Cqlib stores $2^n$ complex amplitudes in a contiguous array. A 64-byte-aligned buffer supports sequential traversal, SIMD loads, and parallel partitioning. Quantum gates update the state vector in place and avoid reallocating the complete state for each operation. With double-precision complex amplitudes, the main state-vector storage is approximately
\begin{equation}
  M_{\mathrm{sv}}(n)=16\times 2^n\ \mathrm{byte}.
\end{equation}

A dense gate acting on a fixed number of target qubits must still access $\mathcal{O}(2^n)$ amplitudes. The Cqlib optimizations do not change the exponential complexity of classical state-vector simulation. Instead, they reduce constant factors through improved data locality, fewer temporary allocations, and fewer instructions.

Cqlib selects an execution path according to the algebraic structure of each gate. An $X$ gate reduces to amplitude swaps. Dedicated component-wise formulas implement $Y$, $Z$, phase, Hadamard, and rotation gates. Only general unitary gates use the generic complex-matrix path. This hierarchy avoids converting every operation into a dense matrix multiplication. It reduces unnecessary complex arithmetic, temporary objects, and memory accesses.

On \texttt{x86\_64} platforms, the kernel detects AVX2 and FMA support at runtime. When available, it enables a SIMD path for the interleaved complex-number layout. Consecutive \texttt{Complex64} amplitudes are mapped to vector registers, and complex multiply-add operations are implemented through multiplication, permutation, addition, and subtraction. The kernel falls back to scalar execution when the required instructions are unavailable. On \texttt{AArch64}, the implementation retains paths that the compiler can vectorize. Cqlib can therefore use vector capabilities without requiring a specific CPU instruction set at deployment time.

State-vector parallelism uses a size-dependent strategy. For small states, the available work does not offset thread creation and scheduling overhead. The current kernel therefore uses 14 qubits as the parallelization threshold. It executes smaller states serially and uses Rayon to update disjoint data blocks for states with 14 or more qubits. This strategy uses multiple cores for large state evolutions without imposing parallel-management overhead on small VQE and QAOA instances or frequently executed unit computations.

For Pauli-string evolution, Cqlib directly implements
\begin{equation}
  U_P(\theta)=\exp\!\left(-\mathrm{i}\theta P/2\right),
\end{equation}
where $P$ is a Hermitian Pauli string. Since $P^2=I$, the operator can be written as
\begin{equation}
  U_P(\theta)=
  \cos\frac{\theta}{2}I-
  \mathrm{i}\sin\frac{\theta}{2}P.
\end{equation}

The native kernel uses bit masks to represent bit flips and phase factors in the Pauli string. For strings that contain only $I/Z$ terms, the evolution applies a phase factor to each amplitude. For strings with $X/Y$ terms, the kernel pairs amplitudes according to the flipped bits and updates them in place. Compared with decomposing the exponential into basis changes, a CX ladder, and an $R_Z$ gate, this fused path reduces intermediate circuit length, instruction dispatches, and repeated state-vector traversals.

\subsubsection{Efficient Support for Variational Quantum Algorithms}

VQE, QAOA, and QML address different problems but share a common execution pattern. They repeatedly bind parameters to a fixed circuit, evolve a quantum state, evaluate an objective function or observable, and update the parameters with a classical optimizer. Cqlib implements these operations through a common parameter representation, state-vector kernel, and operator-execution mechanism. Algorithm frameworks therefore share the same high-performance path.

In VQE, Hamiltonian expectation-value evaluation and ansatz circuits such as UCC and Hamiltonian variational ansatzes repeatedly invoke parameterized rotations and Pauli operators. The centralized parameter representation allows an ansatz structure to be reused across optimization steps. Specialized gate kernels and fused Pauli operations reduce the cost of state preparation and operator application. State evolution, operator evaluation, and result generation remain in the native core. A Python optimizer only submits a parameter vector and receives an energy or expectation value.

In QAOA, the cost and mixer Hamiltonians commonly consist of Pauli terms. Cqlib applies the corresponding exponential operators through the fused Pauli-evolution kernel. This approach reduces auxiliary operations and state-vector traversals introduced by gate-level decomposition. As the circuit depth increases, the parameter table preserves the circuit topology and evaluates only the updated angles. It therefore reduces circuit-reconstruction costs during classical optimization.

Quantum machine-learning workflows often update parameters more frequently and may process batches of samples, deep parameterized circuits, and repeated objective evaluations. Cqlib separates symbolic evaluation from circuit execution, which allows the same quantum model to bind different parameters or encoded data values during training. Coarse-grained language interfaces keep gate-level computation out of Python. They provide a common execution foundation for quantum neural networks, variational classifiers, and quantum-kernel methods.

The Cqlib performance architecture is not limited to local optimization for a single algorithm. It combines a unified circuit representation, centralized parameter management, native state evolution, fused operators, and multilingual interfaces into a reusable computational foundation. VQE, QAOA, and QML retain high-level expressiveness while sharing the safety, consistency, and efficiency of the native implementation. They can also integrate with the Python scientific-computing ecosystem, C/C++ HPC environments, and quantum hardware backends.

\subsection{Quantum Circuit Compilation}

Quantum compilation transforms an algorithm-level logical circuit into an executable circuit that satisfies target-device constraints. A logical circuit describes the algorithmic structure, whereas a physical processor imposes limits on qubit capacity, coupling topology, native gates, and operation errors. 

\begin{figure*}[t]
  \centering
  \includegraphics[width=0.95\textwidth]{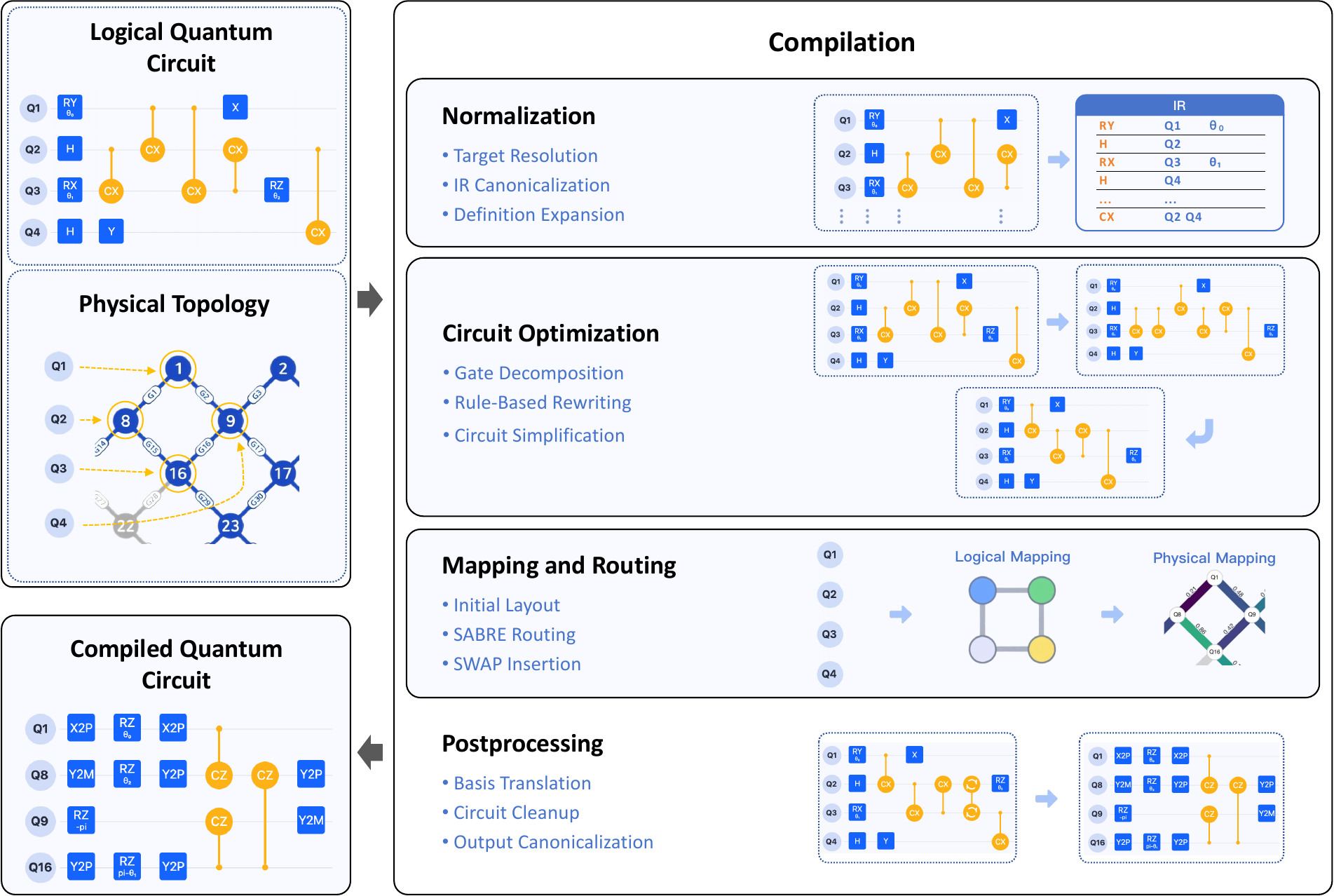}
  \caption{Cqlib quantum compilation workflow}
  \label{fig:quantum-compile}
\end{figure*}

The compiler must preserve circuit semantics while performing circuit normalization, circuit optimization, logical-to-physical qubit mapping, qubit routing, and translation to the target gate set.
Figure~\ref{fig:quantum-compile} shows the overall Cqlib compilation workflow.

Cqlib implements its compiler core in Rust and exposes it through a Python interface. The \texttt{CompilerWorkflow} accepts a standard \texttt{Circuit} and applies normalization, circuit optimization, device adaptation, and postprocessing. It produces a target circuit that operates on physical qubits. The result remains a \texttt{Circuit} object and can be used for simulation, device execution, and result analysis.

The \texttt{CompileConfig} object manages compiler parameters. The \texttt{device}, \texttt{target\_basis}, and \texttt{initial\_layout} fields specify the target device, target gate set, and initial mapping, respectively. Without a device configuration, the compiler performs device-independent decomposition and optimization. With a device configuration, it also checks resources, performs layout and routing, and adapts instructions.

\subsubsection{Normalization}

The circuit-normalization stage resolves compilation targets, checks resource constraints, and standardizes the circuit representation. It reads the device, target gate set, and initial layout from the configuration. It then checks whether the circuit exceeds the device capacity and validates qubit indices, gate parameters, and resource settings. Finally, it converts the input into an intermediate representation shared by subsequent stages.

The compiler expands user-defined gates and composite operations into standard operation sequences. Cqlib also synthesizes general single- and two-qubit unitaries represented by numerical matrices. A single-qubit unitary is converted into a sequence of standard rotations. The Cartan (KAK) decomposition expresses a two-qubit unitary as local single-qubit operations and a nonlocal two-qubit component~\cite{zhang2003geometric}.

For resource-intensive operations such as multi-controlled gates, Cqlib selects a decomposition strategy according to the available ancilla qubits. The \texttt{ResourcePolicy} supports decompositions with no ancilla, clean ancilla, or dirty ancilla. The compiler balances ancilla requirements, basic-gate count, and circuit depth. Circuit normalization produces a standardized circuit with explicit structure and resource information.

\subsubsection{Circuit Optimization}

The circuit-optimization module reduces execution cost through equivalent gate transformations. Cqlib uses a declarative knowledge base to describe equivalent circuit structures. Typical rules cancel inverse gates, merge consecutive rotations, simplify parameters, remove identity structures, and replace composite operations. Rule definitions are independent of the search procedure, which simplifies the addition of new optimization rules and hardware-specific transformations.

The \texttt{KnowledgeRewriter} matches applicable rules within local circuit regions and applies a replacement when it improves the candidate cost. The cost function considers total gate count, two-qubit gate count, circuit depth, and the number of gates outside the target set. Metrics for two-qubit gates receive greater weight because these gates commonly have higher error rates and execution costs.

Cqlib uses the \texttt{CommutationChecker} to identify optimization opportunities between nonadjacent operations. The compiler can reorder commuting operations without changing circuit semantics. This process brings mergeable or canceling gates together and extends the range of local rewrite rules.

Circuit optimization is interleaved with gate decomposition and device adaptation. Optimization before decomposition reduces the expansion of complex operations. Optimization after decomposition removes local redundancies introduced by decomposition templates. In enhanced compilation mode, additional optimization follows routing and target-gate conversion to remove redundancies introduced during device adaptation.

\subsubsection{Mapping and Routing}

The mapping-and-routing module maps a device-independent logical circuit to a specific quantum processor. The \texttt{Device} abstraction describes physical qubits, coupling topology, native gates, and available calibration data. The compiler uses this information for initial layout, circuit routing, and target-gate conversion.

The initial layout defines the starting correspondence between logical and physical qubits, and its quality directly affects routing cost. Cqlib constructs an interaction graph from two-qubit operations in the logical circuit and searches for a mapping on the device coupling graph. It supports Trivial, VF2, and Greedy layout strategies. A user can also provide a mapping through \texttt{initial\_layout}.

The automatic workflow uses SABRE\cite{li2019tackling} for layout and routing by default. When a two-qubit operation violates the physical adjacency constraints, the router generates candidate SWAP operations. It selects an insertion according to the topological cost of the current executable operations and future interactions. By considering both immediate requirements and subsequent circuit structure, SABRE produces an executable circuit while controlling additional SWAP operations and two-qubit depth.

Because heuristic search depends on the initial mapping, Cqlib can run multiple layout and routing trials. It selects a result according to SWAP count, two-qubit gate count, and circuit depth. When calibration data are available, the layout cost can also include two-qubit gate and readout errors. 
Logical qubits that interact frequently can then be mapped to lower-error regions of the device.
Without calibration data, the evaluation uses only the topology and circuit structure.

After routing, the compiler converts the circuit to the gate set specified by \texttt{target\_basis}. If the user does not specify a gate set, the compiler uses the native gates of the device. 
At this stage, operations are ordered to respect circuit dependencies and device connectivity constraints. Device backends handle timing considerations, including gate durations, coherence times, and pulse-level timing optimization.

\subsubsection{Postprocessing}

The postprocessing module removes redundant structures introduced during device adaptation and verifies that the output satisfies the target constraints. Routing, SWAP decomposition, and target-gate conversion can create consecutive rotations, adjacent inverse gates, and other reducible structures. The compiler therefore applies local rule-based optimization after device adaptation. This step reduces circuit cost while preserving the physical mapping.

After optimization, the compiler validates qubit indices, adjacency of two-qubit operations, and consistency with the target gate set. It also verifies that no unresolved composite operations or unsupported instructions remain. The final circuit is returned as a \texttt{Circuit} object together with mapping information such as the initial and final layouts.

The \texttt{WorkflowStepReport} records the status of each compilation stage, changes in circuit size, layout results, and the number of inserted SWAP operations. These reports support diagnosis and comparison of compilation strategies. Cqlib provides \texttt{Normal} and \texttt{Enhanced} compilation modes. \texttt{Normal} uses standard search settings and prioritizes compilation efficiency and stable results. \texttt{Enhanced} expands the search for rule-based optimization, layout, and routing to improve circuit quality at a higher compilation cost.

\section{Benchpress Benchmark Evaluation}
\label{sec:benchpress-benchmark-evaluation}

Benchmarking of quantum computing software is typically performed using representative quantum circuits and processing tasks under unified input conditions, target gate sets, device topologies, and evaluation metrics, so as to comprehensively assess functional coverage, runtime efficiency, and the quality of output circuits. The Benchpress benchmark suite~\cite{nation2025benchpress} is an open benchmarking framework designed for quantum software development kits. It organizes benchmark cases using \texttt{pytest} and collects timing data through \texttt{pytest-benchmark}.

In this work, Cqlib was adapted to the relevant Benchpress benchmark tasks. The evaluation covers circuit construction, compilation for abstract topologies, and compilation for target devices. For the common benchmark cases successfully completed by both Cqlib and Qiskit, we compare circuit construction time, the number and depth of CZ gates in the compiled circuits, and whether the target constraints are satisfied.
We additionally use native Rust and C++ benchmark programs to compare compiler-only runtime across the Rust, C++, and Python interfaces, with identical input circuits, device topologies, target gate sets, and compiler configurations.
The following subsections describe the experimental environment and present the methodologies and results for circuit construction, abstract-topology compilation, device compilation, and multi-language compilation, respectively.

\subsection{Experimental Environment}

The Benchpress benchmarks and cross-language compilation tests reported in this section were run on a macOS computer equipped with an Apple M2 chip.
The system platform identifier was Darwin, and the Python runtime environment was CPython 3.14.5. The experiments used Cqlib and Qiskit 2.5.0. The Rust, C/C++, and Python invocation paths all used the same version of the Cqlib core implementation. The core was compiled in release mode with link-time optimization enabled and \texttt{codegen-units=1}, ensuring that the core compilation configuration remained consistent across the different language interfaces.

\subsection{Evaluation Scope and Methodology}

This section evaluates circuit construction, abstract-topology compilation, device compilation, and multi-language compilation. The first three evaluations compare Cqlib and Qiskit on benchmark cases supported by both SDKs. The circuit construction benchmark reports the median execution time. The abstract-topology and device compilation benchmarks compare the number and depth of CZ gates in the output circuits and report win/tie/loss counts, paired ratios, geometric mean ratios, and aggregate ratios. Both compilation benchmarks use the same target basis gate set, the respective level-2 optimization configurations of the two SDKs, and the same random seed. The output circuits are further checked for compliance with the target gate set, device connectivity constraints, and semantic consistency within the scope of the benchmark. The multi-language compilation benchmark compares the compiler-only runtime of the Rust, C++, and Python interfaces using identical inputs and compilation configurations.

\subsubsection{Circuit Construction}

\begin{figure}[htbp]
    \centering
    \includegraphics[width=\columnwidth]{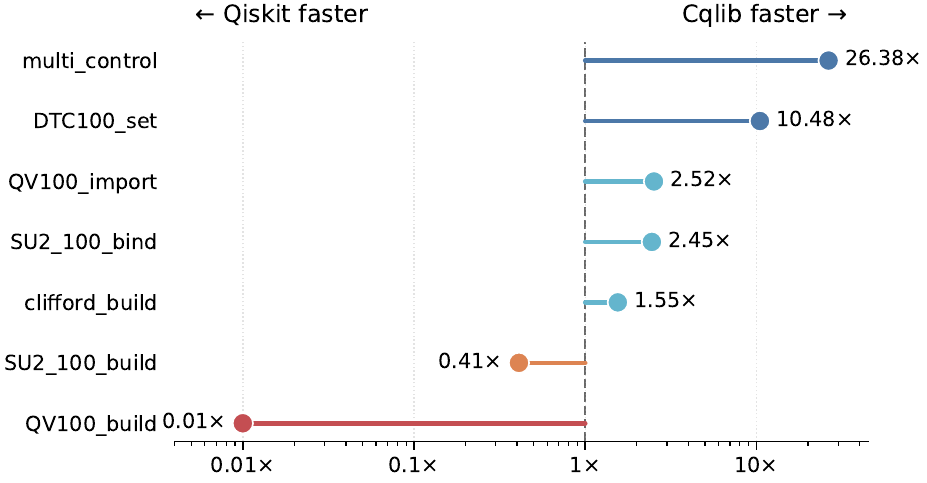}
    \caption{Execution time ratios for circuit construction tasks(Qiskit/Cqlib). Values greater than 1 indicate that Cqlib is faster.}
    \label{fig:circuit_construction_ratio_colored}
\end{figure}

The circuit construction benchmark compares the median execution time of the two SDKs on seven commonly supported tasks. Figure~\ref{fig:circuit_construction_ratio_colored} shows the relative performance, where the execution time ratio is defined as the Qiskit execution time divided by the Cqlib execution time. Therefore, a ratio greater than 1 indicates a lower execution time for Cqlib.

As shown in Figure~\ref{fig:circuit_construction_ratio_colored}, Cqlib achieves lower execution time in five of the seven commonly supported tasks. The most pronounced advantages are observed for multi-controlled gate construction and DTC circuit composition, while smaller advantages are observed for large QASM2 circuit import, parameter binding, and Clifford circuit construction. In contrast, Qiskit shows lower median execution times for the parameterized SU2 and QV100 circuit-construction tasks. The pronounced QV100 difference largely reflects different construction paths: the current Cqlib benchmark implementation relies on repeated Python/SciPy unitary generation and validation, whereas Qiskit uses a dedicated native Quantum Volume builder.

\subsubsection{Abstract-Topology Compilation}

The abstract-topology compilation benchmark uses QASMBench~\cite{li2023qasmbench} circuits as input and evaluates four topology types:
\texttt{all-to-all}, \texttt{square}, \texttt{heavy-hex}, and
\texttt{linear}. The primary statistics are obtained by performing an inner join between the results of the two SDKs based on normalized \texttt{pytest} node IDs. Each topology contains 100 paired benchmark cases using the same OpenQASM inputs.

The compiled outputs are checked using \texttt{circuit\_validator} to confirm that all circuit operations belong to the common target basis gate set and that all two-qubit operations satisfy the connectivity constraints of the corresponding topology. For small- and medium-scale circuits for which exact state-vector simulation is feasible, the output quantum states before and after compilation are further compared. After restoring the logical-qubit ordering, state fidelity insensitive to global phase is calculated. For large-scale circuits where full state-vector simulation is infeasible, correctness verification mainly relies on structural validation, local transformation tests, and compiler unit tests.

In Table~\ref{tab:bp-abstract}, wins, ties, and losses are all reported from the perspective of Cqlib. The geometric mean ratio is the geometric mean of the per-case Cqlib/Qiskit ratios, whereas the aggregate ratio is defined as the total Cqlib metric divided by the total Qiskit metric. For both CZ gate count and CZ depth, a ratio below 1 indicates a lower corresponding metric for Cqlib.

\begin{table}[htbp]
  \centering
  \caption{Compilation results under different abstract topologies}
  \label{tab:bp-abstract}
  \footnotesize
  \setlength{\tabcolsep}{3pt}
  \renewcommand{\arraystretch}{1.08}
  \begin{tabular}{@{}lccc@{}}
    \toprule
    \multicolumn{4}{c}{\textbf{CZ Gate Count}} \\
    \cmidrule(lr){1-4}
    Topology & Win/Tie/Loss & Geometric Mean Ratio & Aggregate Ratio \\
    \midrule
    \texttt{all-to-all} & 6/89/5   & 1.021 & 0.977 \\
    \texttt{square}     & 21/52/27 & 1.017 & 0.959 \\
    \texttt{heavy-hex}  & 28/47/25 & 0.931 & 0.976 \\
    \texttt{linear}     & 23/51/26 & 0.985 & 0.976 \\
    \midrule
    \multicolumn{4}{c}{\textbf{CZ Depth}} \\
    \cmidrule(lr){1-4}
    Topology & Win/Tie/Loss & Geometric Mean Ratio & Aggregate Ratio \\
    \midrule
    \texttt{all-to-all} & 7/86/7   & 0.994 & 0.943 \\
    \texttt{square}     & 30/52/18 & 0.978 & 0.967 \\
    \texttt{heavy-hex}  & 32/48/20 & 0.925 & 1.032 \\
    \texttt{linear}     & 28/51/21 & 0.975 & 1.047 \\
    \bottomrule
  \end{tabular}

  \vspace{0.3em}
  \begin{minipage}{\columnwidth}
    \scriptsize
    Win/Tie/Loss denotes the number of cases in which Cqlib achieves a lower, equal, or higher metric value than Qiskit, respectively.
  \end{minipage}
\end{table}

As shown in Table~\ref{tab:bp-abstract}, among the paired results obtained from identical QASM inputs, relatively noticeable differences are observed for typical samples under the \texttt{heavy-hex} topology. The geometric mean ratios for CZ gate count and CZ depth are 0.931 and 0.925, respectively. However, the aggregate ratio for CZ depth is 1.032, which shows the opposite trend from the geometric mean ratio, indicating that larger benchmark instances have a relatively strong influence on the total CZ-depth metric.

\subsubsection{Device Compilation}

The device compilation benchmark uses the target gate set and coupling constraints of \texttt{FakeTorino}. CZ gate count and CZ depth are compared over 45 paired benchmark cases successfully completed by both SDKs. The compiled outputs are subjected to the same target gate-set, device-connectivity, and semantic-consistency checks used in the abstract-topology compilation benchmark. In Table~\ref{tab:bp-faketorino}, all ratios are defined as the Cqlib metric divided by the corresponding Qiskit metric. Therefore, a ratio below 1 indicates a lower corresponding metric for Cqlib.

\begin{table*}[htbp]
  \centering
  \caption{\texttt{FakeTorino} device compilation results}
  \label{tab:bp-faketorino}
  \footnotesize
  \begin{tabular}{cccccc}
    \toprule
    Metric & No. of Pairs & Win/Tie/Loss
         & Median Paired Ratio & Geometric Mean Ratio & Aggregate Ratio \\
    \midrule
    CZ Gate Count & 45 & 24/11/10 & 0.985 & 0.976 & 0.967 \\
    CZ Depth      & 45 & 24/12/9  & 0.985 & 0.975 & 0.967 \\
    \bottomrule
  \end{tabular}
  
  \vspace{0.3em}
  \begin{minipage}{\textwidth}
  \centering
    \scriptsize
    Win/Tie/Loss is defined from the perspective of Cqlib; lower CZ gate count or depth is considered better.
  \end{minipage}
\end{table*}

As shown in Table~\ref{tab:bp-faketorino}, among the 45 common benchmark cases, Cqlib achieves lower values in 24 cases for both CZ gate count and CZ depth. The median paired ratio is 0.985 for both metrics, while the geometric mean ratios and aggregate ratios are also slightly below 1. 
For the benchmark instances and compilation settings considered here, the circuits produced by Cqlib have slightly lower CZ-gate counts and CZ-gate depths overall than those produced by Qiskit.

\subsubsection{Multi-Language Compilation}

The compilation core of Cqlib is implemented in Rust, with C/C++ and Python interfaces provided through C ABI and PyO3, respectively. The multi-language benchmark uses identical OpenQASM inputs, device coupling edges, target gate sets, compilation modes, and random seeds.
We measure only the runtime of the \texttt{compile} call, excluding process startup, JSON parsing, OpenQASM loading, device construction, and result serialization.
Each of the three invocation paths performs five compilation runs, and the minimum execution time is reported. The results are shown in Table~\ref{tab:bp-3way}.

\begin{table}[htbp]
  \centering
  \caption{Minimum compiler-only runtime across the Rust, C++, and Python interfaces (ms)}
  \label{tab:bp-3way}
  \footnotesize
  \setlength{\tabcolsep}{3.5pt}
  \renewcommand{\arraystretch}{1.08}
  \begin{tabular}{@{}lrrr@{}}
    \toprule
    Benchmark & Rust & C++ & Python \\
    \midrule
    \texttt{deutsch\_n2}       & 24.7  & 23.3  & 22.2 \\
    \texttt{dnn\_n2}           & 237.6 & 258.1 & 252.3 \\
    \texttt{grover\_n2}        & 26.4  & 28.4  & 27.3 \\
    \texttt{ipea\_n2}          & 93.4  & 89.8  & 86.4 \\
    \texttt{iswap\_n2}         & 23.1  & 23.2  & 22.4 \\
    \texttt{quantumwalks\_n2}  & 34.7  & 35.6  & 34.0 \\
    \texttt{basis\_change\_n3} & 84.1  & 81.6  & 78.5 \\
    \texttt{fredkin\_n3}       & 57.9  & 60.2  & 56.8 \\
    \texttt{sat\_n11}          & 746.7 & 797.5 & 826.2 \\
    \texttt{seca\_n11}         & 508.6 & 526.4 & 507.7 \\
    \texttt{cc\_n12}           & 479.6 & 487.6 & 449.9 \\
    \bottomrule
  \end{tabular}
\end{table}

Taking the Rust execution time as the reference in Table~\ref{tab:bp-3way},
the relative difference of the C++ path ranges from $-5.7\%$ to $+8.6\%$,
whereas that of the Python path ranges from $-10.1\%$ to $+10.7\%$.
The fastest language interface varies across benchmark cases. The observed differences are of a magnitude comparable to variations caused by operating-system scheduling, processor frequency fluctuations, memory-allocation state, and short-term system load. All three invocation paths share the same \texttt{Cqlib-core} compilation algorithms and data structures.

The current multi-language comparison covers only the compilation stage and does not yet evaluate end-to-end costs associated with circuit construction, parameter binding, OpenQASM parsing, large-scale matrix transfer, or result-object conversion. Circuit construction tasks in the Benchpress core benchmark, including QV100, DTC100, 100-qubit Clifford circuits, and efficient-SU2 circuits containing 1000 parameters, can be used in future work to extend the multi-language comparison to the complete ``circuit construction plus compilation'' workflow.

\section{Quantum Computing Frameworks Based on Cqlib}
\label{sec:quantum-computing-frameworks-based-on-cqlib}

Quantum algorithms play a central role in quantum computing research and applications. Cqlib provides a general-purpose component library, Cqlib-Algorithm, for building algorithms from reusable components. It also
provides dedicated frameworks for VQE~\cite{peruzzo2014variational},
QAOA~\cite{farhi2014quantum}, and QML~\cite{biamonte2017quantum}, targeting
quantum chemistry, combinatorial optimization, and quantum machine learning, respectively. All four modules build on the high-performance Cqlib Core and use its circuit model, operator representation, and compiler, together with
Cqlib's multi-backend execution system.

\subsection{Cqlib-Algorithm}
\label{subsec:cqlib-algorithm-component-library}
\subsubsection{Positioning and Modular Design}

Cqlib-Algorithm is a general-purpose library for building and running quantum algorithms. Its central design goal is to provide the
fundamental operations needed for common quantum computing tasks, allowing users to reuse them across algorithms and switch between
compatible implementations within the same workflow. To achieve this, the library is organized around protocols, components, and algorithms.

Protocols define common interfaces so that compatible implementations can be used
interchangeably. Components cover tasks such as state preparation, circuit construction, measurement, and classical optimization. Basic components provide individual operations, while composite components combine operations
into reusable subroutines. Algorithms assemble these components into complete workflows for tasks such as search, estimation, and variational optimization. 

\subsubsection{Capability Coverage and Component Composition}

Cqlib-Algorithm covers a broad range of foundational operations used in
standard quantum algorithms, variational methods, and quantum scientific
computing. Its reusable components support the main stages of algorithm
development, from state preparation and circuit construction to
measurement, differentiation, and classical optimization. The library also
provides complete workflows for search and estimation, variational
optimization, quantum dynamics, quantum chemistry and many-body physics,
quantum machine learning, and error mitigation.
Table~\ref{tab:algorithm-component-capabilities} lists representative
capabilities and counts by category.

\begin{table}[t]
\centering
\caption{Capability coverage of Cqlib-Algorithm}
\label{tab:algorithm-component-capabilities}
\footnotesize
\setlength{\tabcolsep}{3pt}
\begin{tabularx}{\columnwidth}{@{}p{0.23\columnwidth}rX@{}}
\toprule
Category & Count & Representative capabilities \\
\midrule
Basic components & 113 &
State preparation and data encoding, parameterized ansatzes and evolution,
oracles and reflections, reversible arithmetic, constraint-preserving mixers,
sampling and measurement, gradients, optimizers, and quantum kernels. \\
Composite components & 16 &
Grover iteration, phase queries, phase-estimation circuits, QSP and QSVT,
Dicke-state components, and circuit transformations for error mitigation. \\
Algorithms and workflows & 43 &
Search, amplitude and phase estimation, variational optimization and quantum
dynamics, quantum chemistry and many-body workflows, quantum machine learning,
and error mitigation. \\
\bottomrule
\end{tabularx}
\end{table}

We demonstrate the composability and interchangeability of algorithm
components using Grover search~\cite{grover1996fast,brassard2000quantum}.
Starting from $\lvert 0^n\rangle$, Grover search prepares an initial
superposition and applies $r$ iterations, yielding
\begin{equation}
\lvert\psi_r\rangle =
\bigl[A(-S_0)A^\dagger S_f\bigr]^r A\lvert 0^n\rangle.
\end{equation}
Here, $A$ prepares the initial superposition, $S_f=I-2\Pi_f$ flips
the phase of the basis states marked by $f(x)=1$, and $\Pi_f$ projects
onto their span. With $S_0=I-2\lvert 0^n\rangle\langle 0^n\rvert$,
the bracketed operator combines phase marking with a reflection about
the prepared state to form one Grover iteration.

Consider a two-qubit search for the marked state $\lvert 10\rangle$.
Figure~\ref{fig:grover-component-composition} shows how Cqlib-Algorithm
assembles the Grover search workflow from reusable components.
\texttt{GroverOperator} combines state preparation with modular
implementations of the zero-state reflection $-S_0$ and the phase
oracle $S_f$ to form one Grover iteration. The resulting workflow
runs through \texttt{grover.run()}, which returns the measurement
counts.

The search predicate and the phase oracle are configured separately,
so either can be changed with a one-line edit. Replacing
\texttt{BooleanFunction} changes which states the algorithm searches
for. Replacing \texttt{PhaseOracle} with a
compatible implementation can change the circuit construction while
preserving the same phase-marking operation.  The same design lets users prototype and validate custom
algorithms by extending base classes such as \texttt{StatePreparation}
and \texttt{Sampler} and combining their own components with existing ones.

\begin{figure}[t]
\centering
\begin{minipage}{\columnwidth}
\begin{lstlisting}[style=cqlibpython,mathescape=true]
from cqlib_algorithms import Grover
from cqlib_algorithms.operators import (
    BooleanFunction, GroverOperator, PhaseOracle,
    StatevectorSampler, UniformSuperposition,
)

# 1. Select the components for $A$ and $S_f$.
preparation = UniformSuperposition(2)
target = BooleanFunction.from_marked_states(2, [2])
oracle = PhaseOracle(target)

# 2. Combine them into $Q = A(-S_0)A^\dagger S_f$.
iteration = GroverOperator(
    state_preparation=preparation, oracle=oracle,
)

# 3. Run $Q^r A\lvert 00\rangle$ and read the counts.
grover = Grover(iteration=iteration, sampler=StatevectorSampler())
result = grover.run(iterations=1, shots=256)
print(result.counts)  # Expected: {'10': 256}
\end{lstlisting}
\end{minipage}

\caption{End-to-end Grover search assembled from reusable
Cqlib-Algorithm components}
\label{fig:grover-component-composition}
\end{figure}

The same approach to composing and replacing components extends beyond
Grover search. Cqlib-Algorithm applies it to quantum query and estimation
algorithms, variational quantum algorithms, and workflows for quantum
dynamics and quantum chemistry. By substituting compatible components, such as different ansatzes or optimizers, researchers can easily adapt these workflows to specific problems and systematically compare alternative designs. Furthermore, this modular design allows users to rapidly prototype and validate novel algorithms by combining existing building blocks.

\subsection{Cqlib-VQE Framework}
\label{subsec:Cqlib-vqe}

\subsubsection{Methodology and Core Capabilities}

VQE is a widely used hybrid quantum--classical method for estimating the ground states of quantum systems~\cite{peruzzo2014variational,mcclean2016theory,kandala2017hardware,tilly2022vqe}. Cqlib-VQE targets ground-state calculations for closed-shell singlet molecules. It integrates electronic-structure modeling, active-space reduction, fermion-to-qubit mapping, unitary coupled-cluster singles and doubles (UCCSD), adaptive operator selection, and energy estimation into a complete quantum chemistry workflow.

The framework uses PySCF\cite{PySCF2018} and OpenFermion\cite{mcclean2020openfermion} to compute molecular integrals, prepare reference states, and construct second-quantized Hamiltonians. Frozen-core and active-space selections control the number of qubits. Hartree--Fock (HF), coupled-cluster singles and doubles (CCSD), full configuration interaction (FCI), and complete active-space configuration interaction (CASCI) calculations use the same effective Hamiltonian and energy offsets as VQE. 
These calculations provide a consistent set of reference energies for assessing the accuracy of the VQE results.

Cqlib-VQE requires a consistent mapping for the Hamiltonian, HF reference state, and excitation generators:
\begin{equation}
  \begin{aligned}
    H_{\mathrm{q}}&=\mathcal{M}(H_{\mathrm{f}}),\\
    G_{j,\mathrm{q}}&=\mathcal{M}(G_{j,\mathrm{f}}),\\
    |\Phi_0^{(\mathcal{M})}\rangle
    &=\mathcal{E}_{\mathcal{M}}
      (\boldsymbol{n}_{\mathrm{HF}}),
  \end{aligned}
\end{equation}

\noindent Here, $H_{\mathrm{f}}$ is the second-quantized fermionic Hamiltonian in the active space, and $H_{\mathrm{q}}$ is its qubit representation. The operators $G_{j,\mathrm{f}}$ and $G_{j,\mathrm{q}}$ are the $j$th excitation generator in the fermionic and qubit representations, respectively. The vector $\boldsymbol{n}_{\mathrm{HF}}$ denotes the HF occupation, $\mathcal{E}_{\mathcal{M}}$ is the corresponding reference-state encoding, and $|\Phi_0^{(\mathcal{M})}\rangle$ is the resulting qubit HF reference state. The framework supports Jordan--Wigner (JW), Bravyi--Kitaev (BK), and parity mappings:
$\mathcal{M}\in\{\mathcal{M}_{\mathrm{JW}},
\mathcal{M}_{\mathrm{BK}},\mathcal{M}_{\mathrm{parity}}\}$.

This mechanism ensures that the Hamiltonian, reference state, and variational operators use the same encoding convention and represent the same fermionic problem.

The framework provides standard UCCSD and an adaptive variational strategy. Standard UCCSD supports CCSD-amplitude initialization and first- or second-order Trotter--Suzuki decomposition. The adaptive strategy iteratively selects important excitations from the UCCSD operator pool and reoptimizes the active parameters. Its objective is to represent electronic correlation with a compact circuit.

For repeated energy evaluations, Cqlib-VQE reuses the Hamiltonian representation, circuit topology, and precompiled Pauli-evolution structures. Native state-vector and Pauli-rotation kernels reduce the cost of parameter iterations. During measurement, qubit-wise commuting (QWC) grouping reduces the number of independent circuits. QCIS compilation and qubit mapping then connect the workflow to Tianyan backends. Local simulation, cloud simulation, and hardware experiments use the same problem definition, energy convention, and result interface.

\subsubsection{Performance Evaluation and Results}

The evaluation uses a single-threaded local state-vector environment under Windows Subsystem for Linux 2 (WSL2). It includes five molecular systems in the STO-3G basis: $\mathrm{H}_2$, $\mathrm{H}_4$, LiH, $\mathrm{BeH}_2$, and $\mathrm{H}_2\mathrm{O}$. The systems range from 4 to 12 qubits. FCI or CASCI energies computed with the same effective active-space Hamiltonian serve as references. The absolute energy error in millihartree (mHa) is
\begin{equation}
  \varepsilon_{\mathrm{mHa}}
  =10^3\left|E_{\mathrm{VQE}}-E_{\mathrm{ref}}\right| ,
\end{equation}

\noindent where $E_{\mathrm{VQE}}$ and $E_{\mathrm{ref}}$ are the variational and exact reference energies in Hartree. Both include the same nuclear-repulsion energy and frozen-core energy offset. The threshold for chemical accuracy is $1.6\,\mathrm{mHa}$. The evaluation has two levels: a matched-workload engine test and an end-to-end workflow test. The first holds the mathematical workload fixed to compare repeated state evolution and expectation-value evaluation. The second retains the high-level quantum chemistry implementation of each framework and evaluates the accuracy and runtime of the complete workflow.

\textbf{Matched-workload engine test.}
All three frameworks use the same initial state, 200 parameterized Pauli rotations, 24 Hamiltonian terms, parameter vector, and qubit count. Probe energies and repeated checks confirm that the numerical tasks are equivalent. This test isolates performance differences in the frequently executed kernels; it does not reproduce the actual number of Pauli terms in each molecular Hamiltonian. Each workload uses 3 warmup evaluations, 30 timed objective evaluations, and 5 repeated measurements. The reported statistic is the median time per evaluation. In Table~\ref{tab:vqe-engine}, the time range gives the minimum and maximum median across the five workloads. The geometric mean relative time is computed from the five runtime ratios with respect to Cqlib-VQE.

\begin{table}[H]
  \centering
  \caption{Matched-workload engine test results}
  \label{tab:vqe-engine}
  \small
  \renewcommand{\arraystretch}{1.12}
  \begin{tabularx}{\columnwidth}{@{}Xcc@{}}
    \toprule
    Framework & \shortstack{Time per evaluation\\range (ms)}
         & \shortstack{Geometric mean\\relative time} \\
    \midrule
    Cqlib & 1.747--6.680 & 1.000 \\
    Qiskit & 140.239--226.016 & 52.331 \\
    PennyLane & 6.133--10.658 & 2.396 \\
    \bottomrule
  \end{tabularx}
\end{table}

In the single-threaded state-vector environment and under the controlled workloads, the median Cqlib-VQE time per evaluation ranges from $1.747$ to $6.680$ ms. The geometric mean runtime ratios of Qiskit and PennyLane relative to Cqlib-VQE are 52.331 and 2.396, respectively. Cqlib-VQE therefore has the lowest measured runtime for repeated objective evaluation under these test conditions. These ratios apply only to the specified environment and mathematical workloads.

\textbf{End-to-end workflow test.}
Each framework runs fixed-UCCSD and adaptive variational workflows, with three repetitions per configuration. The fixed-UCCSD workflow has a budget of 80 objective evaluations, and Cqlib-VQE uses CCSD-amplitude initialization. Each adaptive workflow uses the operator-selection and parameter-optimization strategy of its framework. Because the initialization and adaptive strategies differ, this test compares representative complete workflows rather than identical algorithm configurations. In Table~\ref{tab:vqe-workflow}, the number of accurate systems counts cases whose final error does not exceed the chemical-accuracy threshold. The total time is the sum of the median wall-clock times for the five complete workflows. Relative times use the corresponding Cqlib-VQE workflow as the baseline.

\begin{table}[H]
  \centering
  \caption{End-to-end VQE workflow results for five molecular systems}
  \label{tab:vqe-workflow}
  \small
  \renewcommand{\arraystretch}{1.12}
  \begin{tabularx}{\columnwidth}{@{}Xccc@{}}
    \toprule

    Framework & \shortstack{Accurate\\systems}
         & \shortstack{Total time\\(s)}
         & \shortstack{Relative\\time} \\
    \midrule
    \multicolumn{4}{@{}l}{\textbf{Fixed-UCCSD workflow}} \\
    Cqlib & 3/5 & 16.058 & 1.000 \\
    Qiskit & 3/5 & 230.203 & 14.335 \\
    PennyLane & 2/5 & 149.806 & 9.329 \\
    \midrule
    \multicolumn{4}{@{}l}{\textbf{Adaptive variational workflow}} \\
    Cqlib & 5/5 & 31.795 & 1.000 \\
    Qiskit & 4/5 & 522.745 & 16.441 \\
    PennyLane & 4/5 & 168.083 & 5.286 \\
    \bottomrule
  \end{tabularx}
\end{table}

In the fixed-UCCSD test, Cqlib-VQE reaches chemical accuracy for three systems, with a total median wall-clock time of $16.058$ s. With the adaptive strategy, Cqlib-VQE reaches chemical accuracy for all five systems. The final errors do not exceed $0.368\,\mathrm{mHa}$, and the selected parameter counts range from 1 to 16. 
For the small molecular systems and optimization budgets considered here, the adaptive workflow reaches chemical accuracy for more systems while using only a small number of variational parameters.

The matched-workload test directly compares low-level execution performance. The end-to-end test is also affected by the parameterization, initialization method, operator pool, and adaptive iteration strategy. Its wall-clock and relative times describe complete workflows and should not be interpreted as kernel speedups on identical circuits. Together, the tests evaluate repeated-computation efficiency and quantum chemistry workflows for five systems ranging from 4 to 12 qubits.

\subsection{Cqlib-QAOA Framework}
\label{subsec:Cqlib-qaoa}

\subsubsection{Methodology and Core Capabilities}

QAOA is a widely studied hybrid quantum--classical method for combinatorial optimization~\cite{farhi2014quantum,wang2018quantum,hadfield2019quantum,zhou2020quantum,blekos2024review}. Cqlib-QAOA integrates problem modeling, quadratic unconstrained binary optimization (QUBO)--Ising encoding, cost and mixer circuit construction, classical parameter optimization, quantum sampling, and solution decoding. Effective couplings and feasible-space structures guide circuit construction and optimization. The framework provides an end-to-end workflow for combinatorial optimization.

The framework accepts general QUBO models and provides modeling and decoding functions for MaxCut, the traveling salesperson problem, and vehicle routing. Penalty functions incorporate constraints into the objective, and sampled bit strings are decoded into problem solutions. After normalization and symmetrization, the QUBO coefficients are stored as a sparse Ising model. A single cost layer is decomposed as
\begin{equation}
  U_C(\gamma)
  =e^{-\mathrm{i}\gamma c_0}
   \prod_i R_Z^{(i)}(2\gamma h_i)
   \prod_{i<j}R_{ZZ}^{(i,j)}(2\gamma J_{ij}),
\end{equation}

where $U_C(\gamma)$ is the single-layer cost evolution and $\gamma$ is its evolution parameter. The coefficients $c_0$, $h_i$, and $J_{ij}$ are the energy offset, local field, and pairwise coupling, respectively. The operators $R_Z^{(i)}$ and $R_{ZZ}^{(i,j)}$ are $Z$ rotations on qubit $i$ and qubit pair $(i,j)$. Gates are generated only for nonzero couplings. The circuit size therefore follows the effective problem structure rather than the size of a dense matrix.

Cqlib-QAOA provides an X-mixer for the full binary space and an XY-mixer that preserves the total excitation number. It supports both uniform-superposition and user-defined initial states. A depth-$p$ circuit has $2p$ variational parameters across the cost and mixer layers. Periodic angle constraints control parameter ranges. The framework provides gradient-free optimizers, including simultaneous perturbation stochastic approximation (SPSA)\cite{gacon2021simultaneous}, constrained optimization by linear approximation (COBYLA)\cite{powell2007view}, and Nelder--Mead, for different noise levels and parameter dimensions.

An optimized circuit can run on the local Cqlib state-vector backend or be compiled to QCIS and submitted to a Tianyan backend. The problem model, Ising energy, sampling distribution, and decoded results use consistent semantics. Modeling, optimization, multi--backend experiments, and domain-level analysis can therefore be performed within the same framework.

\subsubsection{Performance Evaluation and Results}

The evaluation uses MaxCut on random weighted undirected graphs to compare the end-to-end state-vector workflows of Cqlib-QAOA, Qiskit, PennyLane, and MindQuantum. For a graph $G=(V,\mathcal{E},\boldsymbol{w})$, let $\mathcal{Z}^{*}$ denote the set of all theoretically optimal bit strings. The cumulative frequency of optimal bit strings measured from the optimized circuit estimates the success probability:
\begin{equation}
\widehat{P}_{\mathrm{succ}} = \frac{1}{N_{\mathrm{shots}}}\sum_{\boldsymbol{z}\in\mathcal{Z}^{*}}N_{\boldsymbol{z}},
\end{equation}

where $V$ and $\mathcal{E}$ are the vertex and edge sets, $\boldsymbol{w}$ contains the edge weights, and $\boldsymbol{z}\in\{0,1\}^{n}$ is a computational-basis bit string. The quantities $N_{\mathrm{shots}}$ and $N_{\boldsymbol{z}}$ are the total number of measurements and the observed count of $\boldsymbol{z}$, respectively.

For every $(n,p)$ configuration, the four frameworks use the same graph instance, state-vector simulation, the SPSA optimizer, and a maximum budget of 100 iterations. The backends are Cqlib Statevector, Qiskit Aer StatevectorSimulator, PennyLane \texttt{default.qubit}, and MindQuantum \texttt{mqmatrix}. The evaluation has two complementary settings. The first fixes $n=6$ and increases the circuit depth from $p=1$ to $p=6$ to study variational expressiveness. The second fixes $p=5$ and increases the graph size from $n=4$ to $n=10$ to study the effect of a larger search space on success probability.

\begin{table}[H]
  \centering
  \caption{Success probability versus circuit depth for fixed $n=6$}
  \label{tab:qaoa-depth}
  \small
  \renewcommand{\arraystretch}{1.10}
  \begin{tabular}{@{}ccccc@{}}
    \toprule
    $p$ & Cqlib & Qiskit & PennyLane & MindQuantum \\
    \midrule
    1 & 0.29 & 0.28 & 0.29 & 0.25 \\
    2 & 0.48 & 0.34 & 0.37 & 0.45 \\
    3 & 0.58 & 0.41 & 0.50 & 0.51 \\
    4 & 0.74 & 0.53 & 0.60 & 0.64 \\
    5 & 0.91 & 0.61 & 0.66 & 0.83 \\
    6 & 0.96 & 0.71 & 0.81 & 0.91 \\
    \bottomrule
  \end{tabular}
\end{table}

Table~\ref{tab:qaoa-depth} shows that the success probability generally increases with $p$ for all four frameworks. Cqlib-QAOA gives the highest value at every test point from $p=2$ to $p=6$. At $p=6$, it reaches 0.96, compared with the highest reference value of 0.91 from MindQuantum. The relative improvement is 5.49\%. Under this test configuration, the Cqlib-QAOA optimization workflow converts the expressiveness of deeper trial states into a higher probability of sampling an optimal solution.

\begin{table}[H]
  \centering
  \caption{Success probability versus problem size for fixed $p=5$}
  \label{tab:qaoa-scale}
  \small
  \renewcommand{\arraystretch}{1.10}
  \begin{tabular}{@{}ccccc@{}}
    \toprule
    $n$ & Cqlib & Qiskit & PennyLane & MindQuantum \\
    \midrule
    4  & 0.99 & 0.92 & 0.97 & 0.99 \\
    5  & 0.94 & 0.84 & 0.88 & 0.85 \\
    6  & 0.91 & 0.69 & 0.66 & 0.83 \\
    7  & 0.85 & 0.60 & 0.56 & 0.77 \\
    8  & 0.72 & 0.56 & 0.44 & 0.56 \\
    9  & 0.67 & 0.53 & 0.52 & 0.58 \\
    10 & 0.47 & 0.30 & 0.24 & 0.39 \\
    \bottomrule
  \end{tabular}
\end{table}

The success probability decreases for all workflows as the number of vertices and the search space increase. Table~\ref{tab:qaoa-scale} shows that Cqlib-QAOA matches the highest reference result at $n=4$ and gives the highest success probability from $n=5$ to $n=10$. For the $n=10$ instance, its success probability is 0.47, a 20.51\% improvement over the highest reference result of 0.39. This trend indicates that Cqlib-QAOA adapts effectively to circuit depth and problem size under the specified optimization budget.

These tests use the high-level QAOA implementation of each framework. Parameter initialization, angle handling, and circuit construction are part of the evaluated workflow. The results therefore compare solution quality for complete algorithms under a common problem and optimization budget; they do not compare simulator speed on identical circuits. They reflect the combined effect of structured initialization, periodic parameter handling, and compact circuit construction without attributing the improvement to any single mechanism.

\subsection{Cqlib-QML Framework}
\label{subsec:Cqlib-qml}

\subsubsection{Methodology and Core Capabilities}

QML is a major research area at the intersection of quantum computing and machine learning~\cite{biamonte2017quantum,schuld2015introduction,cerezo2022challenges}. Cqlib-QML provides composable components for data processing, quantum encoding, parameterized circuits, measurement, classical computation, differentiation, and optimization. Common parameter and gradient interfaces allow model structures, training strategies, and quantum execution methods to be combined independently.

The data module manages classical datasets, while the encoding module transforms the data into quantum data using supported quantum encoding schemes. Encoding circuits, variational circuits, and measurement operators use the common Cqlib quantum-object model. Users can configure circuit depth, rotation gates, entanglement topology, and multiple measurement outputs.

The framework uniformly manages quantum and classical parameters, preserving parameter sharing and allowing modules to be frozen or unfrozen. Quantum and classical components can be flexibly combined in hybrid models.

The differentiation module provides parameter-shift and adjoint methods through the same gradient interface. Parameter-shift differentiation supports backends that return only circuit-execution results. Adjoint differentiation reuses forward state evolution on state-vector simulators to improve gradient efficiency. A common model container coordinates loss functions, classical network layers, optimizers, and learning-rate schedulers. Quantum and classical parameters can be trained jointly or by module.

Based on the proposed Cqlib-QML, five representative QML algorithms have been implemented: Quantum Neural Networks (QNNs)\cite{abbas2021power}, Hybrid Quantum Neural Networks (HQNNs)\cite{long2025hybrid}, Variational Quantum Classifiers (VQCs)\cite{jager2023universal}, Quantum Kernel Methods (QKMs)\cite{thanasilp2024exponential}, and Quantum Support Vector Machines (QSVMs)\cite{li2015experimental}. These implementations demonstrate how different QML models can be constructed and trained within the unified framework. Researchers can further replace or customize data encodings, quantum circuits, measurements, and learning objectives while retaining the same training and execution mechanisms.

\begin{strip}
  \centering
  \captionof{table}{Full-gradient evaluation time for parameterized quantum circuits (s)}
  \label{tab:qml-gradient}
  \footnotesize
  \setlength{\tabcolsep}{4.5pt}
  \renewcommand{\arraystretch}{1.08}
  \begin{tabular}{@{}lcccccccc@{}}
    \toprule
    Framework & \multicolumn{8}{c}{Number of qubits $n$} \\
    \cmidrule(lr){2-9}
    & 2 & 4 & 6 & 8 & 10 & 12 & 14 & 16 \\
    \midrule
    \multicolumn{9}{@{}l}{\textbf{Circuit depth $d=2$}} \\
    Cqlib
      & \textbf{0.000532} & \textbf{0.000709}
      & \textbf{0.001245} & \textbf{0.001725}
      & \textbf{0.002827} & \textbf{0.004881}
      & 0.071446 & 0.251948 \\
    PennyLane
      & 0.008306 & 0.005120 & 0.006251 & 0.007592
      & 0.008950 & 0.011670
      & \textbf{0.019417} & \textbf{0.054215} \\
    Qiskit
      & 0.015819 & 0.031942 & 0.042412 & 0.058639
      & 0.073660 & 0.101429 & 0.175637 & 0.809455 \\
    \midrule
    \multicolumn{9}{@{}l}{\textbf{Circuit depth $d=4$}} \\
    Cqlib
      & \textbf{0.000745} & \textbf{0.001568}
      & \textbf{0.002926} & \textbf{0.004678}
      & \textbf{0.006770} & \textbf{0.012638}
      & 0.141888 & 0.492968 \\
    PennyLane
      & 0.004613 & 0.006203 & 0.010296 & 0.012648
      & 0.015884 & 0.025194
      & \textbf{0.035855} & \textbf{0.101184} \\
    Qiskit
      & 0.027739 & 0.062057 & 0.083927 & 0.118029
      & 0.161403 & 0.215140 & 0.387318 & 1.693029 \\
    \midrule
    \multicolumn{9}{@{}l}{\textbf{Circuit depth $d=6$}} \\
    Cqlib
      & \textbf{0.001108} & \textbf{0.002663}
      & \textbf{0.004811} & \textbf{0.007701}
      & \textbf{0.012813} & \textbf{0.021462}
      & 0.225866 & 0.734512 \\
    PennyLane
      & 0.005847 & 0.009668 & 0.013757 & 0.022604
      & 0.022453 & 0.034469
      & \textbf{0.052626} & \textbf{0.154483} \\
    Qiskit
      & 0.041190 & 0.089370 & 0.132300 & 0.190168
      & 0.261193 & 0.347664 & 0.601619 & 2.633076 \\
    \midrule
    \multicolumn{9}{@{}l}{\textbf{Circuit depth $d=8$}} \\
    Cqlib
      & \textbf{0.001688} & \textbf{0.004811}
      & \textbf{0.007512} & \textbf{0.014354}
      & \textbf{0.020691} & \textbf{0.032749}
      & 0.303794 & 1.011778 \\
    PennyLane
      & 0.007259 & 0.011952 & 0.022313 & 0.022859
      & 0.032970 & 0.043881
      & \textbf{0.075578} & \textbf{0.207841} \\
    Qiskit
      & 0.058748 & 0.130615 & 0.179696 & 0.263575
      & 0.358913 & 0.466632 & 0.834665 & 3.579290 \\
    \bottomrule
  \end{tabular}
\end{strip}

\subsubsection{Performance Evaluation and Results}

A variational quantum model evaluates the gradient of the circuit output with respect to trainable parameters before each update. Gradient-evaluation time therefore directly affects training efficiency. To study the scaling of the Cqlib-QML differentiation module with qubit count and circuit depth, the evaluation constructs functionally equivalent parameterized circuits in Cqlib, PennyLane, and Qiskit. It measures the time required to evaluate the full parameter gradient. Each circuit contains $R_Y$ and $R_Z$ rotations with a linear controlled-NOT (CX) entangling structure. The expectation value of Pauli-$Z$ on qubit 0 provides the scalar output:
\begin{equation}
  \begin{aligned}
    |\psi(\boldsymbol{\theta})\rangle
    &=U(\boldsymbol{\theta})|0\rangle^{\otimes n},\\
    z(\boldsymbol{\theta})
    &=\langle\psi(\boldsymbol{\theta})|
      Z_0|\psi(\boldsymbol{\theta})\rangle,\\
    \boldsymbol{g}(\boldsymbol{\theta})
    &=\nabla_{\boldsymbol{\theta}}z(\boldsymbol{\theta}) .
  \end{aligned}
\end{equation}

Here, $n$ is the number of qubits, $U(\boldsymbol{\theta})$ is a test circuit of depth $d$, and $\boldsymbol{\theta}$ contains all rotation parameters. The state $|\psi(\boldsymbol{\theta})\rangle$ is the circuit output, $Z_0$ is the Pauli-$Z$ operator on qubit 0, and $z(\boldsymbol{\theta})$ is its expectation value. The vector $\boldsymbol{g}(\boldsymbol{\theta})$ is the gradient with respect to all circuit parameters.

The three frameworks use the same circuit structure, parameter count, and observable. Gradient times are measured for 32 combinations of $n\in\{2,4,6,8,10,12,14,16\}$ and $d\in\{2,4,6,8\}$. Cqlib and PennyLane use adjoint differentiation, while Qiskit uses the \texttt{ReverseEstimatorGradient} path. Because each framework uses its own differentiation implementation and state-vector execution path, the results measure their combined efficiency on equivalent gradient tasks. They do not isolate a single gradient kernel. Table~\ref{tab:qml-gradient} reports the measured times. Lower values indicate faster gradient evaluation, and the shortest time for each $(n,d)$ configuration is shown in bold.

The results show distinct behavior across problem sizes. For 2--12 qubits, Cqlib gives the shortest time in all 24 configurations across the four circuit depths. Relative time is defined as the reference-framework time divided by the Cqlib time. Over this range, the geometric mean relative times of PennyLane and Qiskit are 3.163 and 26.148, respectively. Cqlib remains the fastest as the depth increases from 2 to 8, indicating consistent performance across the four tested depths in this qubit range.

For the eight configurations with 14 or 16 qubits, PennyLane is faster than Cqlib, indicating a performance crossover at larger state-vector sizes. Cqlib remains faster than Qiskit in all 32 configurations. The Qiskit-to-Cqlib time ratio ranges from 2.458 to 45.052, with a geometric mean of 15.240. Overall, Cqlib-QML provides a consistent gradient-efficiency advantage for 2--12 qubits and outperforms the Qiskit reverse-gradient path at every tested size. These conclusions apply only to the circuit structure, state-vector backends, and differentiation implementations used in this evaluation. They do not establish a general performance ranking for other configurations.

\section{Cqlib Skills and MCP for Agent-Assisted Quantum Workflows}
\label{sec:cqlib-skills}

\begin{figure*}[t]
  \centering
  \includegraphics[
    width=0.96\textwidth,
  ]{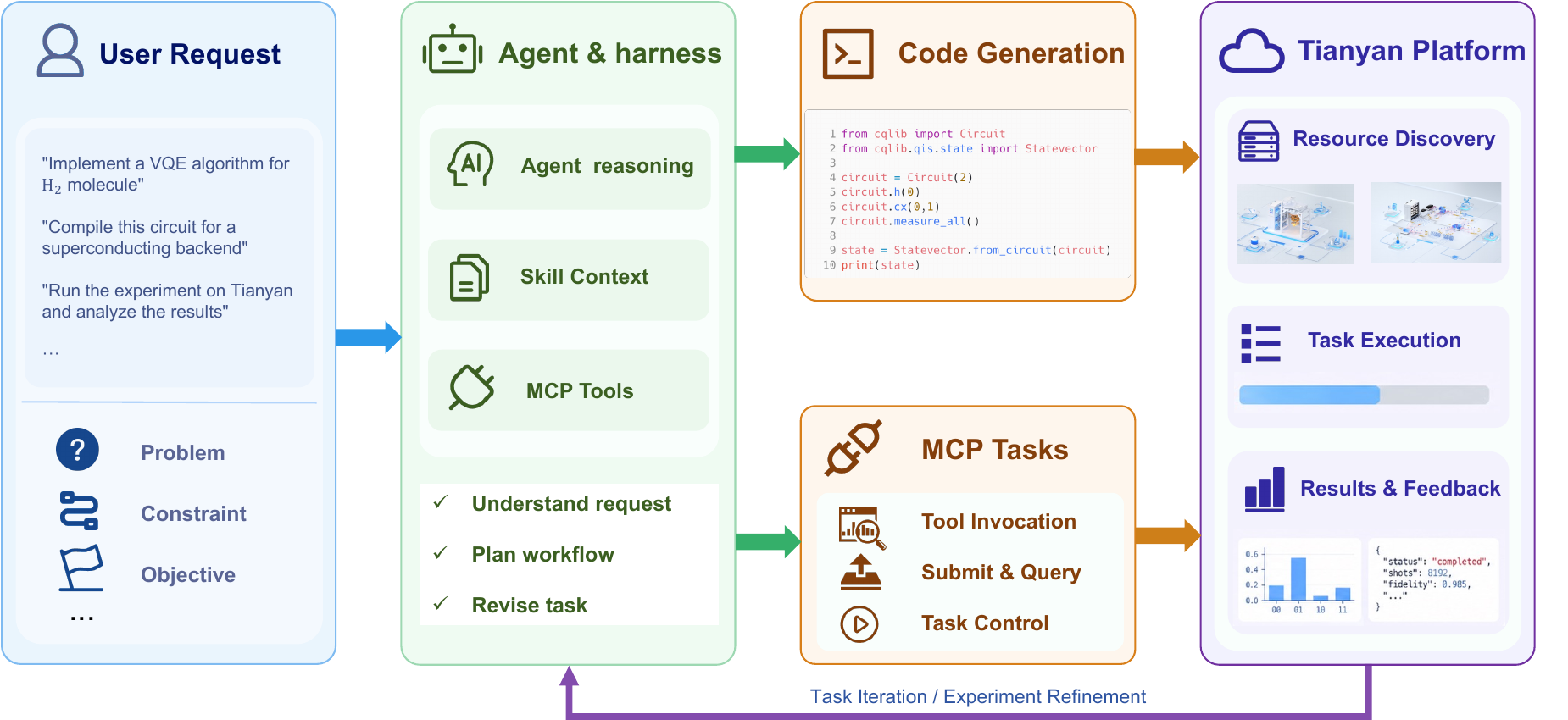}
  \caption{LLM-assisted quantum program generation with Cqlib Skill}
  \label{fig:cqlib-skill-workflow}
\end{figure*}

As large language models and agents become increasingly involved in software development and experimental workflows, Cqlib extends its developer-oriented programming interfaces with Cqlib Skills and services based on the Model Context Protocol (MCP) to support agent-assisted quantum computing workflows \cite{mcp2026specification}. Cqlib Skills provide agents with version-aligned domain knowledge, including APIs, workflows, device constraints, and usage guidelines, while the MCP Server deployed on the Tianyan Quantum Computing Cloud Platform exposes structured tools for backend discovery, controlled experiment submission, expectation-value estimation, and task-status and result retrieval. With these capabilities, an agent can interpret algorithmic and experimental requirements, generate and execute Cqlib programs, or directly invoke cloud quantum-computing services through MCP, and then refine the experiment according to execution and validation results. Figure~\ref{fig:cqlib-skill-workflow}  illustrates this overall workflow. The following subsections describe the knowledge augmentation provided by Cqlib Skills, MCP-based task execution, and the unified agent workflow connecting program generation, backend execution, result validation, and experiment refinement.

\subsection{Cqlib Skills: Source-Grounded Knowledge for Agents}

The knowledge of general-purpose large language models about software libraries is primarily acquired during training and may not reflect the software version currently in use. It may also be insufficient to capture API constraints and execution semantics specific to quantum-computing software. Cqlib Skills organize APIs, common workflows, usage constraints, and examples from the Cqlib ecosystem into domain knowledge that can be loaded on demand. This enables agents to generate programs and plan tasks based on actual software interfaces rather than relying solely on model-internal knowledge. The current Skills cover the Cqlib core library, Cqlib-Tianyan, Cqlib-Pulse, Cqlib-QAOA, and Cqlib-VQE, supporting core quantum-program processing, backend execution, pulse-level workflows, and algorithm-level applications.

Cqlib Skills follow a source-grounded design and remain associated with the corresponding software interfaces, source code, tests, and examples. When interfaces or software versions change, an agent can prioritize the user's current source revision and related implementation details, reducing errors caused by outdated APIs or version mismatches. The Skills do not execute quantum-computing tasks themselves; instead, they form a knowledge layer between the agent and Cqlib, providing task-specific context for subsequent execution through generated Cqlib programs or MCP-based tool calls.

\subsection{MCP-Based Tool and Experiment Execution}

Building on the task-specific knowledge provided by Cqlib Skills, Cqlib exposes structured quantum-computing tools through an MCP Server deployed on the Tianyan Quantum Computing Cloud Platform. The current MCP capabilities cover quantum-backend discovery and inspection, retrieval of existing experiment status and results, controlled submission of single or batched QCIS experiments, and single or batched expectation-value estimation. Experiment-submission operations are separated from read-only queries for backend information, task status, and results, providing a clear boundary between resource inspection and execution.

An agent can either generate and execute a Cqlib program or invoke MCP tools to access cloud computing resources directly. For example, it may first discover an available backend and inspect its configuration, prepare a QCIS program for the selected target, submit the experiment, and then retrieve its status and results using the returned task identifiers. For observable evaluation, the agent can similarly organize measurement tasks and obtain expectation-value results. The MCP Server therefore forms the execution layer between the agent and the Tianyan platform in Figure~\ref{fig:cqlib-skill-workflow}, complementing Cqlib’s program-generation and compilation capabilities with structured tools for backend discovery, experiment submission, and result retrieval. The underlying Tianyan platform already provides the corresponding mechanisms for device queries, task submission, status tracking, and result retrieval.

\subsection{Unified Agent Workflow and Validation}

Cqlib Skills and MCP together form a unified quantum-computing workflow for agents. As illustrated in Figure~\ref{fig:cqlib-skill-workflow}, an agent first interprets the algorithmic, device, and experimental requirements specified by the user and obtains task-relevant software knowledge from Cqlib Skills.

It can then follow one of two execution paths: the first generates a complete Cqlib program for local circuit construction, simulation, compilation, and result processing; the second, by contrast, invokes structured MCP tools to access backend discovery, experiment submission, and result retrieval on the Tianyan platform.
Both paths reuse Cqlib’s common quantum-program representation, compilation, and execution mechanisms and eventually converge at result analysis and validation.

Validation results are used not only to determine whether the program or experiment satisfies the target constraints, but also to guide subsequent result-guided experiment refinement. Based on compilation information, execution results, and validation outcomes, an agent can revise the circuit, compiler configuration, target backend, or experimental parameters. Local generation and validation can be iterated directly, whereas cloud experiments require verification of the existing task state before a new task is submitted within the user-authorized experiment scope. In the current MCP deployment, Cqlib Skills provide task knowledge, MCP provides structured execution interfaces, and the Tianyan platform serves as the execution backend for the actual computation and experiments. Together, these components connect requirement interpretation, program generation or tool invocation, execution, validation, and experiment refinement.

\section{Related Work}
\label{sec:related-work}

The scope of quantum software frameworks continues to expand. General-purpose development kits such as Qiskit and ProjectQ integrate circuit construction, simulation, compilation, and backend access within a common programming environment~\cite{javadiabhari2024qiskit,steiger2018projectq}. As quantum applications target physical devices, software stacks must also address performance, device constraints, integration with classical computation, and task execution. Related work therefore spans high-performance simulation and circuit processing, compilation for target devices, hybrid quantum--classical programming, algorithm components, and backend execution services.

Performance is an important objective in quantum software. Qulacs provides efficient quantum circuit simulation, while QuEST supports large simulations on multicore CPUs, GPUs, and distributed systems~\cite{suzuki2021qulacs,jones2019quest}. In a general-purpose quantum software stack, performance-critical operations include not only state evolution but also circuit construction, parameter binding, and compiler transformations. A common design is therefore to implement these operations in a native kernel and expose them through high-level interfaces. Cqlib implements circuit processing, quantum information computation, and compilation in Rust. Its Python and C interfaces use the same underlying implementation.

Quantum compilers increasingly account for constraints imposed by target devices. Frameworks such as Qiskit and t|ket$\rangle$ organize gate decomposition, circuit optimization, initial layout, routing, and target-gate conversion into backend-oriented compilation workflows~\cite{javadiabhari2024qiskit,sivarajah2020tket}. A logical circuit is not a directly executable program; it must be adapted to the available qubits, coupling topology, and native gates of a device. Cqlib provides a staged compiler based on its common circuit representation. It also supports conversion among QCIS, OpenQASM 2.0, and OpenQASM 3.0, connecting circuit construction, compiler optimization, and hardware execution within the same software stack.

Quantum algorithm development has extended software frameworks toward hybrid quantum--classical workflows. Variational algorithms such as VQE and QAOA combine problem modeling, parameterized circuits, classical optimization, backend execution, and result analysis~\cite{peruzzo2014variational,farhi2014quantum}. PennyLane connects quantum nodes to automatic differentiation and machine-learning toolchains~\cite{bergholm2018pennylane}. In this setting, an algorithm component is a problem-oriented application unit rather than only a circuit template. Cqlib-VQE, Cqlib-QAOA, and Cqlib-QML target quantum chemistry, combinatorial optimization, and QML, respectively. They reuse the circuit, compilation, and execution capabilities of Cqlib and demonstrate the use of a common software foundation across application domains. Cqlib-Algorithm complements these problem-oriented frameworks by providing reusable quantum operators and composable components for developing new algorithms and customizing hybrid quantum–classical workflows.

Access to physical devices and cloud platforms further expands the responsibilities of quantum software. Authentication, backend discovery, device information, task submission, status polling, result retrieval, and error handling depend on a specific platform. These functions require a clear boundary from general circuit and compiler capabilities. Cqlib constructs, analyzes, compiles, and converts quantum programs, while \texttt{Cqlib-Tianyan} communicates with the Tianyan platform. QCIS and common result structures connect the two components. Together, they provide a workflow from local algorithm development and device adaptation to backend execution. Cqlib is not intended to replace a specialized simulator, compiler, or algorithm framework. It connects multilingual high-performance development, compilation for target devices, algorithm applications, and hardware execution through a common circuit representation.

High-performance simulation is another foundation of quantum software. Qulacs combines a native circuit simulator with numerical optimizations for research workloads~\cite{suzuki2021qulacs}. QuEST supports large quantum-state simulations on multicore CPUs, multiple GPUs, and distributed systems~\cite{jones2019quest}. In addition to dense state-vector methods that store all $2^n$ amplitudes, sparse-state methods reduce the number of amplitudes maintained during evolution~\cite{jaques2022leveraging,sun2026sparqsim}. Sparse methods can reduce time and memory costs when a state remains sufficiently sparse, but their applicable conditions differ from those of dense state-vector simulation. This prior work primarily addresses state representations, performance optimization, and low-level simulator implementations. Cqlib instead uses a Rust core for performance-sensitive circuit and quantum information operations within a broader software stack. Python and C interfaces support both algorithm development and integration with existing systems. The Benchpress evaluation and multilingual tests in this paper measure circuit-processing and compilation performance under common workloads.

Prior work has established mature methods for general quantum programming, differentiable hybrid computation, retargetable compilation, and high-performance simulation. Cqlib does not aim to replace these frameworks with a single specialized module. Instead, it combines a native Rust kernel, Python and C interfaces, QCIS compatibility, device-aware compilation, and general backend execution interfaces around a common quantum program representation, with Cqlib-Tianyan serving as one concrete backend implementation.
This paper does not assume that Cqlib outperforms existing frameworks in execution speed or compilation quality. The Benchpress and multilingual experiments evaluate Cqlib under the reported workloads and conditions.

\section{Conclusion and Outlook}
\label{sec:conclusion-and-outlook}

This paper presents Cqlib, a high-performance quantum software stack built around a unified quantum program representation and a native computational core. Cqlib integrates quantum program construction, quantum-information processing, simulation, device-aware compilation, and backend execution under consistent program semantics. Its Rust core, Python and C interfaces, compiler pipeline, and unified result-processing mechanisms provide a reusable software foundation for quantum algorithm development and deployment. Benchpress and multilingual compilation tests evaluate its functional coverage, compilation correctness, and performance across language interfaces. The Cqlib-Algorithm provides reusable components and customizable workflows for quantum algorithm development, while the Cqlib-VQE, Cqlib-QAOA, and Cqlib-QML frameworks demonstrate the extensibility of this common software foundation to quantum chemistry, combinatorial optimization, and quantum machine learning, respectively.

As quantum processors improve in scale, gate fidelity, dynamic control, and quantum--classical integration, quantum software will need to support larger and more complex workloads that depend closely on hardware characteristics. Future development of Cqlib will expand support for dynamic circuits and hybrid quantum–classical execution, refine device abstractions and improve compilation across hardware platforms, and continue to optimize high-performance simulation and computational primitives.

For fault-tolerant quantum computing, Cqlib will explore software abstractions for logical circuits, quantum error correction, and resource estimation.

In addition, Cqlib provides Cqlib Skills and Model Context Protocol (MCP) services to support program construction, compilation, simulation, and backend execution, enabling AI coding agents to assist quantum programming and experiment orchestration through natural-language-driven workflows. We expect these agent-oriented interfaces to lower the barrier to quantum software development and to support increasingly automated quantum experiments on real hardware. These capabilities extend the Cqlib software stack toward increasingly automated development, compilation, and execution of quantum programs across simulation and hardware backends.

\subsection*{\textit{\textbf{}{Tianyan}} Quantum Group}

\noindent Hantao Sun$^{1*}$, Shengbin Wang$^{1*}$, Jianjian Gao$^{1}$, Zhen Wang$^1$, Zilong Zha$^1$, 
Jiawang Wan$^{1}$, Yue Tian$^{1}$, Lijun Sun$^{1}$, Hanze Guo$^{1}$, Tianqi Zong$^{1}$, Zhaowei Li$^{1}$,  Yin Kan$^{1}$, 
Yuchen He$^{1}$, Yu Fang$^{1}$, Mingzheng Zhu$^{1}$, Taiping Sun$^{1}$, Hantao Zhang$^{1}$, Hongyu Chen$^{1}$, Tong Gao$^{1}$, 
Xinfang Zhang$^{1}$, Hanyi Wang$^1$, Qiankun Wang$^1$, Jiasheng Hu$^1$, Junhe Wang$^1$, Bihao Guo$^1$, Chunwang Liu$^1$, Yang Li$^1$, Yu Fan$^1$, Rui Xu$^1$, Yuanhang Hu$^1$, Rui Yang$^1$, Weizhi Tao$^1$,
Jie Chen$^1$, Lei Miao$^1$, Jiping Fu$^1$, Wenya Huang$^1$, Mian Tao$^1$, Wei Liu$^1$, Zirun Luo$^1$, Mengtian Zhou$^1$, Yuting Zhang$^1$, He Huang$^1$, Yang Jin$^1$, Yuan Liu$^1$, Jingen Wang$^1$, 
Shuyue Zheng$^1$, Jian Ding$^1$, Canxia Zhou$^1$, Fan Zhang$^1$, Shulin Yu$^1$, Liangfang Cui$^1$, Xu Han$^1$, Yuejia Wang$^1$, 
Hao Li(李浩)$^1$,
Xixi Zhang$^1$, Xin Ma$^1$, 
Junpeng Chen$^{1}$, Aoyu Zhang$^{1}$, Shunxin Cao$^{1}$, 
Pei Wan$^1$, Zhizhou Liang$^1$, Muyu Liang$^1$,
Chuanxin Huang$^1$, Songyuan Chen$^1$, Mingrui Wei$^1$, Fenghui Wang$^1$, Xueqian Long$^1$, Chi Zhang$^1$, Biying Wang$^1$, Zhaoting Li$^1$, Jun Li$^1$, Hao Li(李昊)$^1$.

\noindent $^1$ China Telecom Quantum Information Technology Group Co., Ltd., Hefei 230031, China\\

\noindent $^*$ Corresponding authors (Hantao Sun: sunhantao@chinatelecom.cn, Shengbin Wang: wangshengbin@chinatelecom.cn)

\bibliographystyle{unsrt}
\bibliography{main}          

\begin{thebibliography}{10}

\bibitem{preskill2018nisq}
John Preskill.
\newblock Quantum computing in the {NISQ} era and beyond.
\newblock {\em Quantum}, 2:79, 2018.

\bibitem{cerezo2021vqa}
Marco Cerezo, Andrew Arrasmith, Ryan Babbush, Simon~C. Benjamin, Suguru Endo, Keisuke Fujii, Jarrod~R. McClean, Kosuke Mitarai, Xiao Yuan, Lukasz Cincio, and Patrick~J. Coles.
\newblock Variational quantum algorithms.
\newblock {\em Nature Reviews Physics}, 3(9):625--644, 2021.

\bibitem{javadiabhari2024qiskit}
Ali Javadi-Abhari, Matthew Treinish, Kevin Krsulich, Christopher~J. Wood, Jake Lishman, Julien Gacon, Simon Martiel, Paul~D. Nation, Lev~S. Bishop, Andrew~W. Cross, Blake~R. Johnson, and Jay~M. Gambetta.
\newblock Quantum computing with {Qiskit}.
\newblock {\em arXiv preprint arXiv:2405.08810}, 2024.

\bibitem{cross2017openqasm}
Andrew~W. Cross, Lev~S. Bishop, John~A. Smolin, and Jay~M. Gambetta.
\newblock Open quantum assembly language.
\newblock {\em arXiv preprint arXiv:1707.03429}, 2017.

\bibitem{cross2022openqasm3}
Andrew~W. Cross, Ali Javadi-Abhari, Thomas Alexander, Niel de~Beaudrap, Lev~S. Bishop, Steven Heidel, Colm~A. Ryan, John Smolin, Jay~M. Gambetta, and Blake~R. Johnson.
\newblock Openqasm 3: A broader and deeper quantum assembly language.
\newblock {\em ACM Transactions on Quantum Computing}, 3(3):1--50, 2022.

\bibitem{nation2025benchpress}
Paul~D. Nation, Abdullah~Ash Saki, Sebastian Brandhofer, Luciano Bello, Shelly Garion, Matthew Treinish, and Ali Javadi-Abhari.
\newblock Benchmarking the performance of quantum computing software for quantum circuit creation, manipulation and compilation.
\newblock {\em Nature Computational Science}, 5(5):427--435, 2025.

\bibitem{tianyan2025cloud}
{Tianyan Quantum Group}.
\newblock Tianyan: Cloud services with quantum advantage.
\newblock {\em arXiv preprint arXiv:2512.10504}, 2025.

\bibitem{wu2021strong}
Yulin Wu, Wan-Su Bao, Sirui Cao, Fusheng Chen, Ming-Cheng Chen, Xiawei Chen, Tung-Hsun Chung, Hui Deng, Yajie Du, Daojin Fan, Ming Gong, Cheng Guo, Chu Guo, Shaojun Guo, Lianchen Han, Linyin Hong, He-Liang Huang, Yong-Heng Huo, Liping Li, Na~Li, Shaowei Li, Yuan Li, Futian Liang, Chun Lin, Jin Lin, Haoran Qian, Dan Qiao, Hao Rong, Hong Su, Lihua Sun, Liangyuan Wang, Shiyu Wang, Dachao Wu, Yu~Xu, Kai Yan, Weifeng Yang, Yang Yang, Yangsen Ye, Jianghan Yin, Chong Ying, Jiale Yu, Chen Zha, Cha Zhang, Haibin Zhang, Kaili Zhang, Yiming Zhang, Han Zhao, Youwei Zhao, Liang Zhou, Qingling Zhu, Chao-Yang Lu, Cheng-Zhi Peng, Xiaobo Zhu, and Jian-Wei Pan.
\newblock Strong quantum computational advantage using a superconducting quantum processor.
\newblock {\em Physical Review Letters}, 127(18):180501, 2021.

\bibitem{zhang2003geometric}
Jun Zhang, Jiri Vala, Shankar Sastry, and K.~Birgitta Whaley.
\newblock Geometric theory of nonlocal two-qubit operations.
\newblock {\em Physical Review A}, 67(4):042313, 2003.

\bibitem{li2019tackling}
Gushu Li, Yufei Ding, and Yuan Xie.
\newblock Tackling the qubit mapping problem for nisq-era quantum devices.
\newblock In {\em Proceedings of the 24th International Conference on Architectural Support for Programming Languages and Operating Systems}, pages 1001--1014. Association for Computing Machinery, 2019.

\bibitem{li2023qasmbench}
Ang Li, Samuel Stein, Sriram Krishnamoorthy, and James Ang.
\newblock {QASMBench}: A low-level quantum benchmark suite for {NISQ} evaluation and simulation.
\newblock {\em ACM Transactions on Quantum Computing}, 4(2), 2023.

\bibitem{peruzzo2014variational}
Alberto Peruzzo, Jarrod McClean, Peter Shadbolt, Man-Hong Yung, Xiao-Qi Zhou, Peter~J Love, Al{\'a}n Aspuru-Guzik, and Jeremy~L O’brien.
\newblock A variational eigenvalue solver on a photonic quantum processor.
\newblock {\em Nature communications}, 5(1):4213, 2014.

\bibitem{farhi2014quantum}
Edward Farhi, Jeffrey Goldstone, and Sam Gutmann.
\newblock A quantum approximate optimization algorithm.
\newblock {\em arXiv preprint arXiv:1411.4028}, 2014.

\bibitem{biamonte2017quantum}
Jacob Biamonte, Peter Wittek, Nicola Pancotti, Patrick Rebentrost, Nathan Wiebe, and Seth Lloyd.
\newblock Quantum machine learning.
\newblock {\em Nature}, 549(7671):195--202, 2017.

\bibitem{grover1996fast}
Lov~K Grover.
\newblock A fast quantum mechanical algorithm for database search.
\newblock In {\em Proceedings of the twenty-eighth annual ACM symposium on Theory of computing}, pages 212--219, 1996.

\bibitem{brassard2000quantum}
Gilles Brassard, Peter Hoyer, Michele Mosca, and Alain Tapp.
\newblock Quantum amplitude amplification and estimation.
\newblock {\em arXiv preprint quant-ph/0005055}, 2000.

\bibitem{mcclean2016theory}
Jarrod~R. McClean, Jonathan Romero, Ryan Babbush, and Al{\'a}n Aspuru-Guzik.
\newblock The theory of variational hybrid quantum-classical algorithms.
\newblock {\em New Journal of Physics}, 18(2):023023, 2016.

\bibitem{kandala2017hardware}
Abhinav Kandala, Antonio Mezzacapo, Kristan Temme, Maika Takita, Markus Brink, Jerry~M. Chow, and Jay~M. Gambetta.
\newblock Hardware-efficient variational quantum eigensolver for small molecules and quantum magnets.
\newblock {\em Nature}, 549(7671):242--246, 2017.

\bibitem{tilly2022vqe}
Jules Tilly, Hongxiang Chen, Shuxiang Cao, Dario Picozzi, Kanav Setia, Ying Li, Edward Grant, Leonard Wossnig, Ivan Rungger, George~H. Booth, and Jonathan Tennyson.
\newblock The variational quantum eigensolver: A review of methods and best practices.
\newblock {\em Physics Reports}, 986:1--128, 2022.

\bibitem{PySCF2018}
Qiming Sun, Timothy~C. Berkelbach, Nick~S. Blunt, George~H. Booth, Sheng Guo, Zhendong Li, Junzi Liu, James~D. McClain, Elvira~R. Sayfutyarova, Sandeep Sharma, Sebastian Wouters, and Garnet Kin-Lic Chan.
\newblock Pyscf: the python-based simulations of chemistry framework.
\newblock {\em WIREs Computational Molecular Science}, 8(1):1340, 2018.

\bibitem{mcclean2020openfermion}
Jarrod~R McClean, Nicholas~C Rubin, Kevin~J Sung, Ian~D Kivlichan, Xavier Bonet-Monroig, Yudong Cao, Chengyu Dai, E~Schuyler Fried, Craig Gidney, Brendan Gimby, et~al.
\newblock Openfermion: the electronic structure package for quantum computers.
\newblock {\em Quantum Science and Technology}, 5(3):034014, 2020.

\bibitem{wang2018quantum}
Zhihui Wang, Stuart Hadfield, Zhang Jiang, and Eleanor~G. Rieffel.
\newblock Quantum approximate optimization algorithm for maxcut: A fermionic view.
\newblock {\em Physical Review A}, 97(2):022304, 2018.

\bibitem{hadfield2019quantum}
Stuart Hadfield, Zhihui Wang, Bryan O'Gorman, Eleanor~G. Rieffel, Davide Venturelli, and Rupak Biswas.
\newblock From the quantum approximate optimization algorithm to a quantum alternating operator ansatz.
\newblock {\em Algorithms}, 12(2):34, 2019.

\bibitem{zhou2020quantum}
Leo Zhou, Sheng-Tao Wang, Soonwon Choi, Hannes Pichler, and Mikhail~D. Lukin.
\newblock Quantum approximate optimization algorithm: Performance, mechanism, and implementation on near-term devices.
\newblock {\em Physical Review X}, 10(2):021067, 2020.

\bibitem{blekos2024review}
Kostas Blekos, Dean Brand, Andrea Ceschini, Chiao-Hui Chou, Rui-Hao Li, Komal Pandya, and Alessandro Summer.
\newblock A review on quantum approximate optimization algorithm and its variants.
\newblock {\em Physics Reports}, 1068:1--66, 2024.

\bibitem{gacon2021simultaneous}
Julien Gacon, Christa Zoufal, Giuseppe Carleo, and Stefan Woerner.
\newblock Simultaneous perturbation stochastic approximation of the quantum fisher information.
\newblock {\em Quantum}, 5:567, 2021.

\bibitem{powell2007view}
Michael~JD Powell.
\newblock A view of algorithms for optimization without derivatives.
\newblock {\em Mathematics Today-Bulletin of the Institute of Mathematics and its Applications}, 43(5):170--174, 2007.

\bibitem{schuld2015introduction}
Maria Schuld, Ilya Sinayskiy, and Francesco Petruccione.
\newblock An introduction to quantum machine learning.
\newblock {\em Contemporary Physics}, 56(2):172--185, 2015.

\bibitem{cerezo2022challenges}
Marco Cerezo, Guillaume Verdon, Hsin-Yuan Huang, Lukasz Cincio, and Patrick~J Coles.
\newblock Challenges and opportunities in quantum machine learning.
\newblock {\em Nature computational science}, 2(9):567--576, 2022.

\bibitem{abbas2021power}
Amira Abbas, David Sutter, Christa Zoufal, Aur{\'e}lien Lucchi, Alessio Figalli, and Stefan Woerner.
\newblock The power of quantum neural networks.
\newblock {\em Nature computational science}, 1(6):403--409, 2021.

\bibitem{long2025hybrid}
Changzhou Long, Meng Huang, Xiucai Ye, Yasunori Futamura, and Tetsuya Sakurai.
\newblock Hybrid quantum-classical-quantum convolutional neural networks.
\newblock {\em Scientific Reports}, 15(1):31780, 2025.

\bibitem{jager2023universal}
Jonas J{\"a}ger and Roman~V Krems.
\newblock Universal expressiveness of variational quantum classifiers and quantum kernels for support vector machines.
\newblock {\em Nature Communications}, 14(1):576, 2023.

\bibitem{thanasilp2024exponential}
Supanut Thanasilp, Samson Wang, Marco Cerezo, and Zo{\"e} Holmes.
\newblock Exponential concentration in quantum kernel methods.
\newblock {\em Nature communications}, 15(1):5200, 2024.

\bibitem{li2015experimental}
Zhaokai Li, Xiaomei Liu, Nanyang Xu, and Jiangfeng Du.
\newblock Experimental realization of a quantum support vector machine.
\newblock {\em Physical review letters}, 114(14):140504, 2015.

\bibitem{mcp2026specification}
{Model Context Protocol Contributors}.
\newblock Model context protocol specification, 2026.
\newblock Specification version 2026-07-28.

\bibitem{steiger2018projectq}
Damian~S. Steiger, Thomas H{\"a}ner, and Matthias Troyer.
\newblock Projectq: An open source software framework for quantum computing.
\newblock {\em Quantum}, 2:49, 2018.

\bibitem{suzuki2021qulacs}
Yasunari Suzuki, Yoshiaki Kawase, Yuya Masumura, Yuria Hiraga, Masahiro Nakadai, Jiabao Chen, Ken~M. Nakanishi, Kosuke Mitarai, Ryosuke Imai, Shiro Tamiya, Takahiro Yamamoto, Tennin Yan, Toru Kawakubo, Yuya~O. Nakagawa, Yohei Ibe, Youyuan Zhang, Hirotsugu Yamashita, Hikaru Yoshimura, Akihiro Hayashi, and Keisuke Fujii.
\newblock Qulacs: A fast and versatile quantum circuit simulator for research purpose.
\newblock {\em Quantum}, 5:559, 2021.

\bibitem{jones2019quest}
Tyson Jones, Anna Brown, Ian Bush, and Simon~C. Benjamin.
\newblock {QuEST} and high performance simulation of quantum computers.
\newblock {\em Scientific Reports}, 9:10736, 2019.

\bibitem{sivarajah2020tket}
Seyon Sivarajah, Silas Dilkes, Alexander Cowtan, Will Simmons, Alec Edgington, and Ross Duncan.
\newblock t|ket$\rangle$: A retargetable compiler for {NISQ} devices.
\newblock {\em Quantum Science and Technology}, 6(1):014003, 2021.

\bibitem{bergholm2018pennylane}
Ville Bergholm, Josh Izaac, Maria Schuld, Christian Gogolin, Carsten Blank, Keri McKiernan, and Nathan Killoran.
\newblock Pennylane: Automatic differentiation of hybrid quantum-classical computations.
\newblock {\em arXiv preprint arXiv:1811.04968}, 2018.

\bibitem{jaques2022leveraging}
Samuel Jaques and Thomas H{\"a}ner.
\newblock Leveraging state sparsity for more efficient quantum simulations.
\newblock {\em ACM Transactions on Quantum Computing}, 3(3), 2022.

\bibitem{sun2026sparqsim}
Tai-Ping Sun, Yun-Jie Wang, Zhao-Yun Chen, Cheng Xue, Huan-Yu Liu, Xi-Ning Zhuang, Xiao-Fan Xu, Yu-Chun Wu, and Guo-Ping Guo.
\newblock {SparQSim}: An efficient quantum algorithm simulator leveraging state sparsity.
\newblock {\em Chinese Physics B}, 2026.

\end{thebibliography}

\appendix

\titleformat{\section}[display]
  {\normalfont\Large\bfseries\centering}
  {Appendix \thesection}
  {0.5em}
  {}

\section{Cqlib Software Ecosystem}

\begin{figure*}[t]
\centering
\includegraphics[width=0.8\textwidth]{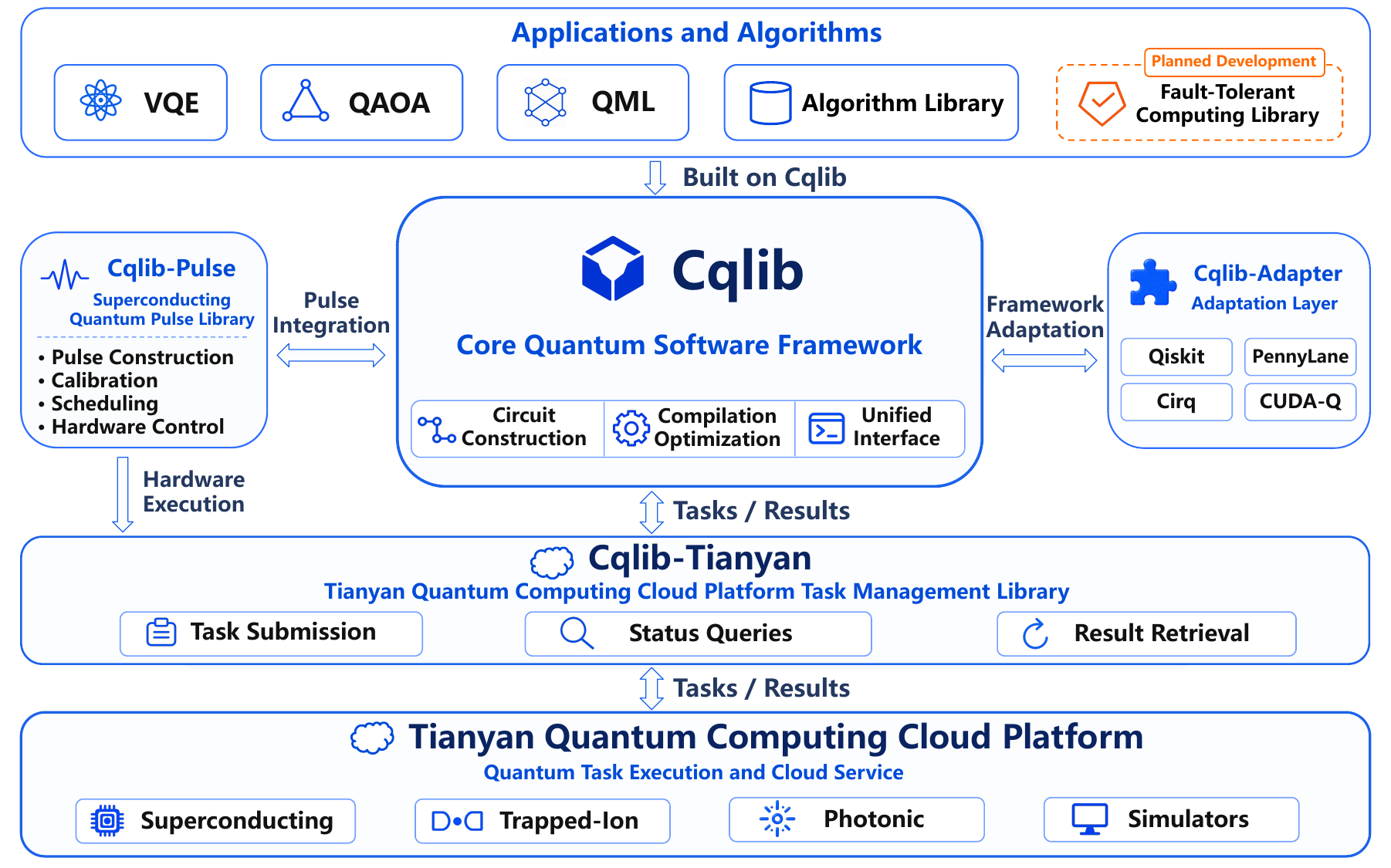}
\caption{Architecture and components of the Cqlib software ecosystem}
\label{fig}
\end{figure*}

The Cqlib software ecosystem is built upon a unified quantum program representation and a native computing core. It connects application and algorithm development, superconducting quantum pulse processing, external quantum computing framework adaptation, and quantum cloud platform execution, forming a complete software architecture covering quantum program construction, simulation and compilation, platform task submission, status monitoring, and result retrieval. As shown in Fig.~\ref{fig}, the ecosystem is centered around Cqlib and includes four existing frameworks built on Cqlib, namely Algorithm, VQE, QAOA, and QML, the superconducting quantum pulse library Cqlib-Pulse, the interoperability layer Cqlib-Adapter, the Tianyan quantum cloud platform task management library Cqlib-Tianyan, and the Tianyan quantum cloud platform. In addition, a fault-tolerant computing library is planned for future development.

Cqlib serves as the core component of the software ecosystem. It provides fundamental representations of quantum circuits, parameters, quantum gates, quantum operators, and device models, while supporting general-purpose tasks including circuit construction, quantum state simulation, compilation optimization, format conversion, and result processing. In the figure, its core capabilities are summarized as quantum circuit construction, compilation optimization, and a unified execution interface. The unified program representation and computing interface provide a common software foundation for algorithm frameworks, pulse libraries, adapters, and platform task management libraries within the ecosystem.

The application and algorithm layer consists of four existing modules. The Cqlib-Algorithm organizes public protocols, reusable operators, composite components, and complete algorithms. Cqlib-VQE, Cqlib-QAOA, and Cqlib-QML are parallel domain-oriented frameworks for quantum chemistry, combinatorial optimization, and quantum machine learning.  All four frameworks are built upon Cqlib and can utilize the circuit construction, parameter management, quantum state simulation, compilation, and execution interfaces provided by Cqlib to organize corresponding quantum algorithm workflows. Each framework maintains its own algorithmic structure and application interfaces, while delegating common quantum program processing tasks to Cqlib.

Cqlib-Pulse is a pulse library designed for superconducting quantum computers. It mainly handles pulse construction, calibration, scheduling, and hardware control-related tasks, involving device-specific objects such as waveforms, control channels, and timing sequences. The library provides pulse-level implementations corresponding to quantum operations. As an integrated component of the Cqlib software ecosystem, Cqlib-Pulse can connect with the hardware execution workflow of the Tianyan quantum cloud platform through Cqlib-Tianyan, enabling experiments on real superconducting quantum devices.

The fault-tolerant computing library shown in the figure is a planned future component.
Building on the quantum operator representation and computational capabilities already available in Cqlib, the planned operator library is intended to provide reusable interfaces for operator construction, combination, transformation, and computation.
The fault-tolerant computing library is intended to provide software components for quantum error correction codes, logical quantum operations, and fault-tolerant computing workflows.

Cqlib-Adapter provides an interoperability layer between Cqlib and external quantum software frameworks.
It currently supports Qiskit 2.x, Cirq 1.7+, PennyLane 0.45.x, and CUDA-Q 0.15.x on non-Windows platforms. 
The adapter converts quantum circuits or quantum kernels from external frameworks into Cqlib’s unified program representation. The conversion covers qubits, quantum gates, parameters, and measurement operations.
Cqlib can then analyze and simulate the converted programs, optimize and compile them for target devices, and generate QCIS instructions.
After execution, the adapter organizes and returns results according to the original framework's measurement semantics, qubit ordering, and result formats. 
This allows users to retain their existing framework APIs and workflows without switching to Cqlib’s native APIs or writing QCIS instructions manually.

Cqlib-Tianyan and the Tianyan quantum cloud platform together provide cloud-based task management and execution within the Cqlib ecosystem. Cqlib-Tianyan handles authentication, queries backend and device information, submits tasks, monitors their status, and retrieves results.
It interfaces with Cqlib through task descriptions and structured execution results. The Tianyan quantum cloud platform receives submitted tasks and schedules corresponding quantum computing or simulation resources. In the figure, superconducting, trapped-ion, photonic, and simulation platforms represent different types of connected computing resources. Through this execution channel, quantum programs generated by Cqlib and device-level programs provided by the pulse library can enter the corresponding execution workflow, while execution results are returned to upper-layer frameworks for further processing.

\section{Quantum Control Instruction Set (QCIS)}
The Quantum Control Instruction Set (QCIS) is a text-based instruction set for controlling superconducting quantum computing hardware\cite{wu2021strong}. In Cqlib, QCIS serves as an interface between the internal circuit representation and QCIS-compatible execution backends. A user first constructs a logical quantum circuit from qubits, classical bits, parameters, and quantum operations. The compiler then performs gate decomposition, qubit mapping, routing, and target-gate conversion according to the physical qubits, coupling topology, and native gate set of the selected device. Finally, the resulting physical circuit is serialized into QCIS instructions.

\subsection{QCIS Representation and Cqlib Support}

A QCIS program consists of text instructions in execution order, with one quantum operation per line. In the circuit-level Cqlib implementation, each instruction contains an opcode, target qubits, and optional parameters. Its general form is
\begin{center}
\texttt{OPCODE QUBIT [QUBIT ...] [PARAMETER ...]}
\end{center}
Here, \texttt{OPCODE} identifies the operation, \texttt{QUBIT} specifies a target qubit, and \texttt{PARAMETER} provides a value for a parameterized operation. A qubit is written as an uppercase \texttt{Q} followed by a nonnegative integer index, as in \texttt{X2P Q0} and \texttt{CZ Q0 Q1}.

Cqlib supports fixed single-qubit gates, parameterized gates, $\pm\pi/2$ rotations, and multiqubit gates in QCIS. The \texttt{X2P}, \texttt{X2M}, \texttt{Y2P}, and \texttt{Y2M} instructions denote $\pm\pi/2$ rotations about the $x$ and $y$ axes:
\begin{equation}
\begin{aligned}
X2P &= R_x\left(\frac{\pi}{2}\right), &
X2M &= R_x\left(-\frac{\pi}{2}\right),\\
Y2P &= R_y\left(\frac{\pi}{2}\right), &
Y2M &= R_y\left(-\frac{\pi}{2}\right),
\end{aligned}
\label{eq:qcis-half-pi-gates}
\end{equation}
where $R_{\alpha}(\theta)=\exp[-\mathrm{i}\theta\sigma_{\alpha}/2]$ and $\sigma_{\alpha}$ is a Pauli operator. Cqlib also supports parameterized half-$\pi$ rotations about an arbitrary axis $\boldsymbol{n}(\phi)=(\cos\phi,\sin\phi,0)$ in the $xy$ plane:
\begin{equation}
\begin{aligned}
XY2P(\phi)
&=\exp\left[-\mathrm{i}\frac{\pi}{4}
\left(\cos\phi\,\sigma_x+\sin\phi\,\sigma_y\right)\right],\\
XY2M(\phi)
&=\exp\left[+\mathrm{i}\frac{\pi}{4}
\left(\cos\phi\,\sigma_x+\sin\phi\,\sigma_y\right)\right].
\end{aligned}
\label{eq:qcis-xy2-gates}
\end{equation}
Table~\ref{tab:qcis-supported-operations} summarizes the circuit-level QCIS gates supported by Cqlib.

\begin{table*}[t]
\centering
\caption{Circuit-level QCIS quantum gates supported by Cqlib}
\label{tab:qcis-supported-operations}
\begin{tabular}{p{0.22\textwidth}p{0.72\textwidth}}
\hline
Category & Instructions \\
\hline
Fixed single-qubit gates
& \texttt{X}, \texttt{Y}, \texttt{Z}, \texttt{H}, \texttt{S},
\texttt{SD}, \texttt{T}, \texttt{TD} \\
Parameterized gates
& \texttt{RX}, \texttt{RY}, \texttt{RZ}, \texttt{RXY}, \texttt{U},
\texttt{XY}, \texttt{PHASE} \\
Half-$\pi$ rotations
& \texttt{X2P}, \texttt{X2M}, \texttt{Y2P}, \texttt{Y2M},
\texttt{XY2P}, \texttt{XY2M} \\
Multiqubit gates
& \texttt{CX}, \texttt{CY}, \texttt{CZ}, \texttt{CCX}, \texttt{SWAP},
\texttt{CRX}, \texttt{CRY}, \texttt{CRZ}, \texttt{RXX}, \texttt{RYY},
\texttt{RZZ}, \texttt{RZX}, \texttt{FSIM} \\
\hline
\end{tabular}
\end{table*}

The Rust core implements QCIS parsing and generation, and the Python interface exposes both operations. In the \texttt{Cqlib.ir.qcis} module, \texttt{dumps} serializes a \texttt{Circuit} object as QCIS text, while \texttt{loads} parses QCIS text into a \texttt{Circuit}. The parsed circuit can then be inspected, analyzed, or recompiled. The conversion reports errors for unsupported operations and invalid parameters rather than silently dropping operations or changing circuit semantics.

\subsection{QCIS Conversion for Device Execution}

QCIS generation follows the Cqlib device-compilation workflow. Given a logical circuit $C$, the compiler performs gate decomposition, logical-to-physical qubit mapping, circuit routing, and target-gate conversion for the selected device. These steps produce a physical circuit $C'$. Let $Q_{\mathrm{P}}$ be the physical-qubit set, $E$ the coupling-edge set, and $\mathcal{G}_{\mathrm{native}}$ the native gate set. Before QCIS generation, the physical circuit must satisfy
\begin{equation}
Q(C')\subseteq Q_{\mathrm{P}},
\qquad
G(C')\subseteq \mathcal{G}_{\mathrm{native}}.
\end{equation}

For every two-qubit gate $g(q_i,q_j)$ in the circuit, the physical qubits must also satisfy the coupling constraint of the target device. If $E$ denotes the physical edges that support two-qubit operations, then
\begin{equation}
(q_i,q_j)\in E.
\end{equation}

When several mappings or routes are feasible, the compiler can evaluate candidates with device-calibration data, including two-qubit gate errors, readout errors, and coupling-edge availability. After device adaptation, the QCIS module converts the physical circuit into text instructions.
The generated QCIS program can then be submitted through Cqlib-Tianyan for execution on the selected backend.

\end{document}